\documentclass[referee]{aa}

\usepackage{natbib}
\usepackage{graphicx}

\usepackage{color}
\definecolor{shamrockgreen}{rgb}{0.0, 0.62, 0.38}
\usepackage{bm}

\newcommand{\beq}{\begin{equation}}
\newcommand{\eeq}{\end{equation}}
\newcommand{\beqa}{\begin{eqnarray}}
\newcommand{\eeqa}{\end{eqnarray}}

\usepackage{color}           

\begin{document}

\title{Spectroscopic Observations and Modelling of Impulsive Alfv\'{e}n Waves Along a Polar Coronal Jet}

\author{P.~Jel\'\i nek\inst{1}, A.K. Srivastava\inst{2}, K.~Murawski\inst{3}, P.~Kayshap\inst{4}, B.N.~Dwivedi\inst{2}}

\institute{University of South Bohemia, Faculty of Science, Institute of Physics and Biophysics, Brani\v sovsk\'a 10, CZ -- 370 05 \v{C}esk\'e Bud\v{e}jovice, Czech Republic 
\email{pjelinek@prf.jcu.cz} 
\and
Department of Physics, Indian Institute of Technology (Banaras Hindu University), Varanasi-221005, India 
\and
Group of Astrophysics, UMCS, ul. Radziszewskiego 10, 20-031 Lublin, Poland
\and
Inter University Centre for Astronomy and Astrophysics, Ganeshkhind, Pune, India
}
\date{Received / Accepted }

\abstract
{The magnetic reconnection in the solar corona results in {\textbf an} impulsively generated Alfv\'en waves which drive a polar jet.}
{Using the Hinode/EIS 2$"$ spectroscopic observations, we study the intensity, velocity, and FWHM variations of the strongest Fe XII 195.12 \AA\ line along the jet to find the signature of Alfv\'en waves. We simulate numerically the impulsively generated Alfv\'en waves within the vertical Harris current-sheet, forming the jet plasma flows, and mimicking their observational signatures.}
{Using the FLASH code and the atmospheric model with embedded weakly expanding magnetic field configuration within a vertical Harris current-sheet, we solve the two and half-dimensional (2.5-D) ideal magnetohydrodynamic (MHD) equations to study the evolution of Alfv\'en waves and vertical flows forming the plasma jet.}
{At a height of $\sim 5~\mathrm{Mm}$ from the base of the jet, the red-shifted velocity component of Fe XII 195.12 \AA\ line attains its maximum ($5~\mathrm{km\,s}^{-1}$) which converts into a blue-shifted one between the altitude of $5-10~\mathrm{Mm}$. The spectral intensity continously increases up to $10~\mathrm{Mm}$, while FWHM still exhibits the low values with almost constant trend. This indicates that the reconnection point within the jet's magnetic field topology lies in the corona $5-10~\mathrm{Mm}$ from its footpoint anchored in the Sun's surface. Beyond this height, FWHM shows a growing trend. This may be the signature of Alfv\'en waves that impulsively evolve due to reconnection and propagate along the jet. From our numerical data, we evaluate space- and time- averaged Alfv\'en waves velocity amplitudes at different heights in the jet's current-sheet, which contribute to the non-thermal motions and spectral line broadening. The synthetic width of Fe XII $195.12~\mathrm{\AA}$ line exhibits similar trend of increment as in the observational data, possibly proving the existence of impulsively generated (by reconnection) Alfv\'en waves which propagate along the jet.}
{The numerical simulations show that the impulsive perturbations in the transversal component of velocity at the reconnection point can excite the Alfv\'en waves. These waves can power the plasma jet higher into the polar coronal hole as vertical plasma flows are also associated with these waves due to pondermotive force. The simulated Alfv\'en waves match well with the observed non-thermal broadening along the jet, which may provide direct spectroscopic evidence of the impulsively excited Alfv\'en waves within the polar jet.}


\titlerunning{Alfv\'en waves driven polar jet}

\authorrunning{P.~Jel\'\i nek et al.}
\maketitle

\section{Introduction}

Polar coronal jets are well observed, large-scale, and confined plasma transients in the solar atmosphere (see Nistic{\`o} et al. 2009; Nistic{\`o} \&  Zimbardo 2012). They can significantly contribute to the energy transport and the formation of nascent supersonic wind. The exact driving mechanisms of such jets are still debated. Broadly speaking, the two candidates, namely magnetic reconnection and magnetohydrodynamic (MHD) waves, are known to trigger these jets (e.g., Shibata 1982; Yokoyama \& Shibata 1995; Cirtain et al. 2007; Nishizuka et al. 2008; Filippov et al. 2009; Pariat et al. 2009, 2015, 2015; Srivastava \& Murawski 2011; Kayshap et al. 2013, and references cited therein). The observations of Alfv\'en waves associated with the polar coronal jets are either related to the imaging observations of photospherically driven these waves propagating along the jet (Cirtain et al. 2007) or some transversal perturbations  evolved during the reconnection at 
the base of the jet (Nishizuka et al. 2008). {\bf Such Alfv\'en waves are also ubiquitous in the localized magnetic structures (e.g., spicules, prominences, small-scale chromospheric fluxtubes) as well as in the large-scale corona (e.g., De Pontieu et al. 2007; Okamoto et al. 2007; Tomczyk et al. 2007, Jess et al. 2009; Mathioudakis et al. 2013, and references cited therein)}. 

\begin{figure*}
\centering
\includegraphics[width=18.0cm]{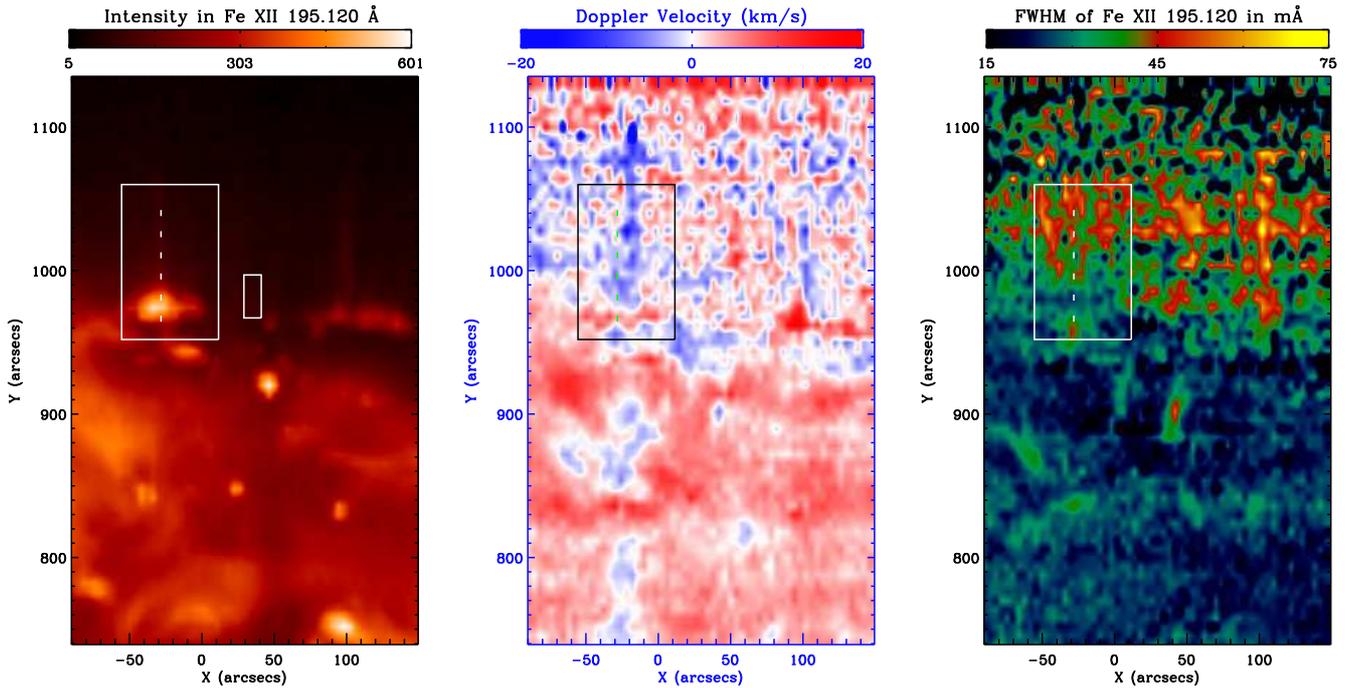}


\caption{The intensity, Doppler velocity, and FWHM maps of Fe XII $195.12~\mathrm{\AA}$ line, showing the polar jet
moving off the limb outward. 
The white line is the position of the slit along the jet upon which the spectral parameters are estimated. The footpoint of the jet where slit's lowest part lies, has the coordinates of ($\mathbf{-28"}$, $\mathbf{964"}$). The white rectangular box is chosen at the limb of north polar corona from where the integrated line-profile of Fe XII 195.12~$\mathrm{\AA}$ is derived and fitted in order to obtain its rest wavelength.}
\label{fig304}
\end{figure*}

Detection of Alfv\'en wave in solar magnetic structures is not yet well established as these waves are incompressible. In the polar coronal plasma, these waves are observed in the form of spectral-line profile variations (Banerjee et al. 1998, Harrison et al. 2002). Narrowing of the spectral line-width is attributed to the dissipation of small-amplitude Alfv\'en waves (Harrison et al. 2002; O'Shea et al. 2005; Bempord \& Abbo 2012; Dwivedi et al. 2014, and references cited therein). The growth and dissipation of Alfv\'en waves have been modelled in the corona, which also explain the observed line-width variations (e.g., Pekenulu et al. 2002; Dwivedi \& Srivastava 2006; Chmielewski et al. 2013, 2014). In particular, Chmielewski et al. (2013) have reported that the observed line broadening, particularly in the polar corona, can be explained in terms of impulsively generated non-linear Alfv\'en waves. {\bf Non}-linear Alfv\'en waves {\bf may be the} likely candidates for transporting energy in the {\bf solar corona (Murawski et al. 2015a).} These waves can also power the large-scale, as well as confined plasma transients, e.g., solar jets, nascent wind{\bf , and plasma flows (Murawski et al., 2015b)}.

It has been reported that the Alfv\'en waves power various jets (e.g., X-ray jets, spicules) and exist in a variety of coronal magnetic structures. However, they are identified as kink waves (Van Doorsselaere et al. 2008a, 2008b).
The existence of mixed radial and azimuthal waves are also reported in theory and observations (Goossens et al. 2009, 2012; Tian et al. 2012; Srivastava \& Goossens 2013; and references cited therein). {\bf Numerical simulations support the general interpretation of the observed oscillations as a coupling of the kink and Alfv\'en waves (see e.g. Pascoe et al. 2010, 2011).}
Kamio et al. (2010) have reported the spectroscopic observations of a rotating coronal jet and interpreted it as an evolution of kink waves and instability. The {\bf solar jets} are found to be driven by {\bf various physical processes}, e.g., the direct magnetic reconnection generated Lorentz force (Nishizuka et al. 2008), the reconnection generated pulse (Srivastava \& Murawski 2011), emergence and internal reconnection in small-scale kinked flux-tubes (Kayshap et al. 2013), etc. {\bf However,} the pure Alfv\'en waves driven coronal jets are difficult to detect due to  observational constraints and the dynamic nature of the jet's typical magnetic field and plasma configuration. 

{\bf In the present paper, we study the Hinode/EIS spectroscopic observations of a polar jet. We find the spectroscopic signatures of impulsively generated Alfv\'en waves and associated plasma flows within the jet. We model the observed physical processes (waves and flows) in the jet as a natural consequence of reconnection generated velocity pulse in the vertical and gravitationally stratified Harris current-sheet lying in the appropriate model atmosphere with VAL-III~C temperature (Vernazza et al. 1981).} The structure of the present paper is as follows. In Sect. 2, we present the observational results. Sect. 3 describes numerical model, governing equations, initial conditions, perturbations, and numerical solutions. In Sect. 4, the numerical results are shown. Discussion and conclusions are outlined in the last section.

\begin{figure*}
\centering
\mbox{
\includegraphics[width=6.0cm]{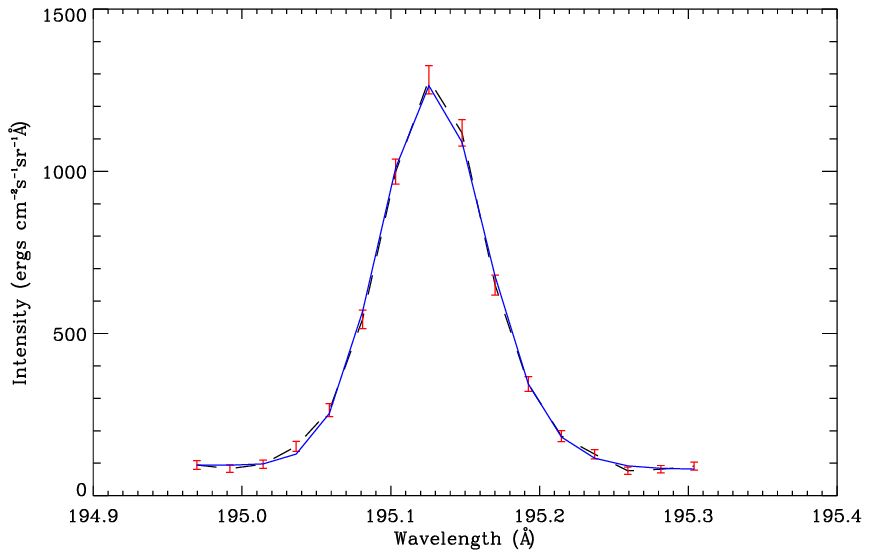}
\includegraphics[width=6.0cm]{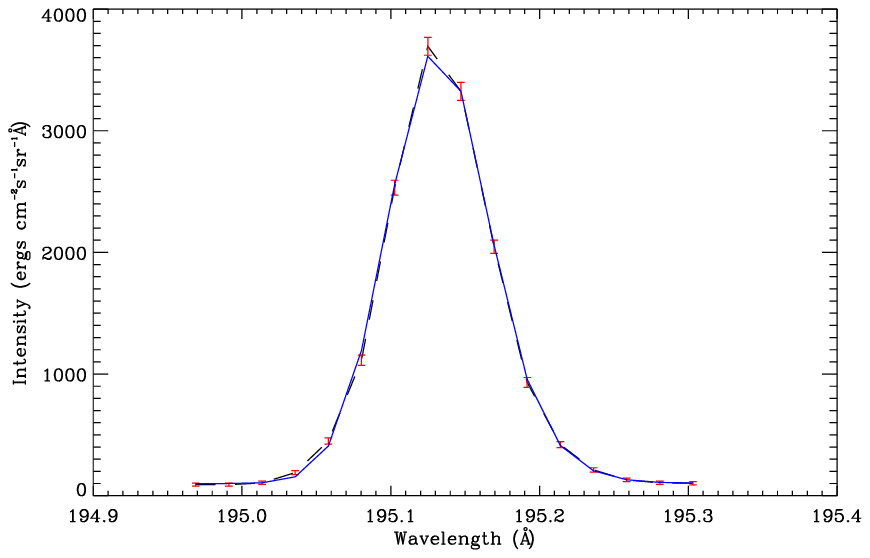}
\includegraphics[width=6.0cm]{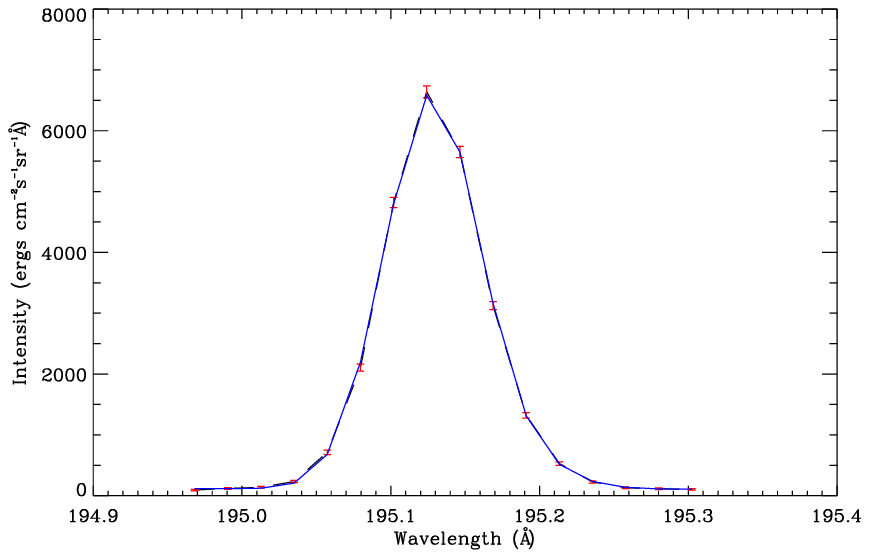}
}
\mbox{
\includegraphics[width=6.0cm]{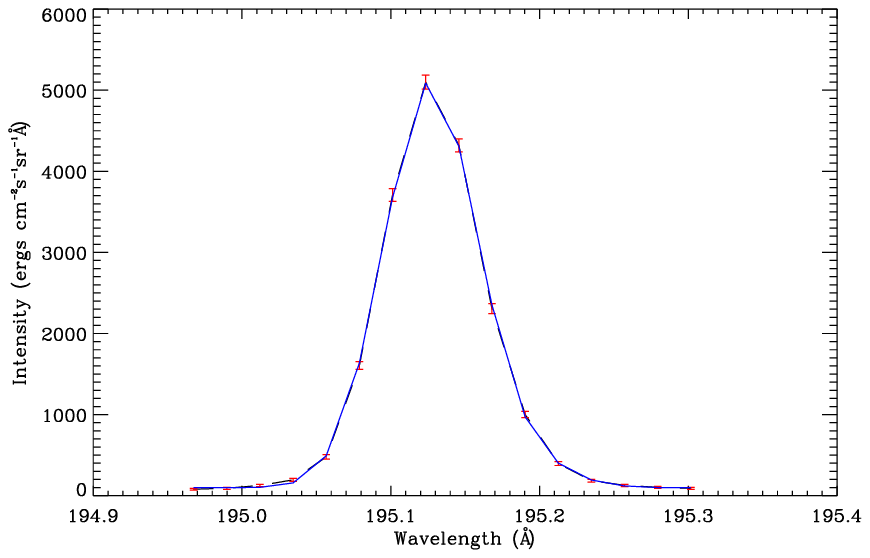}
\includegraphics[width=6.0cm]{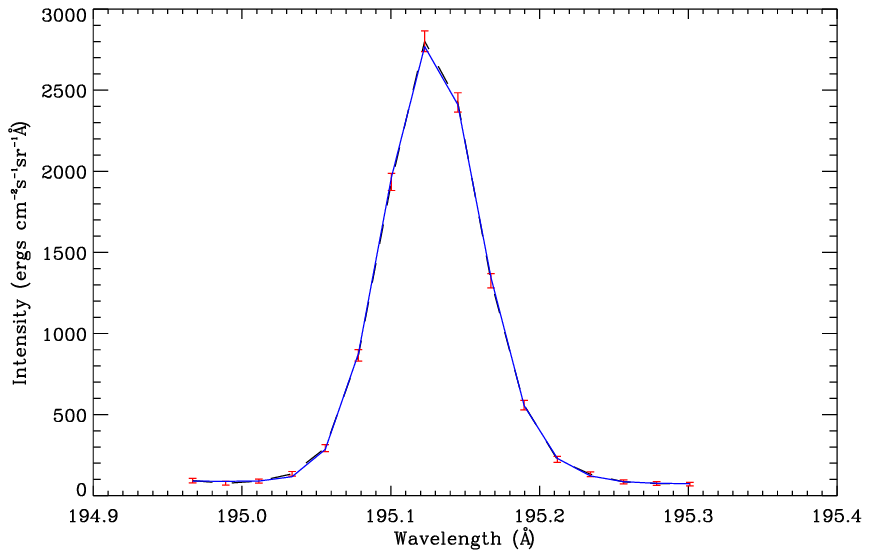}
\includegraphics[width=6.0cm]{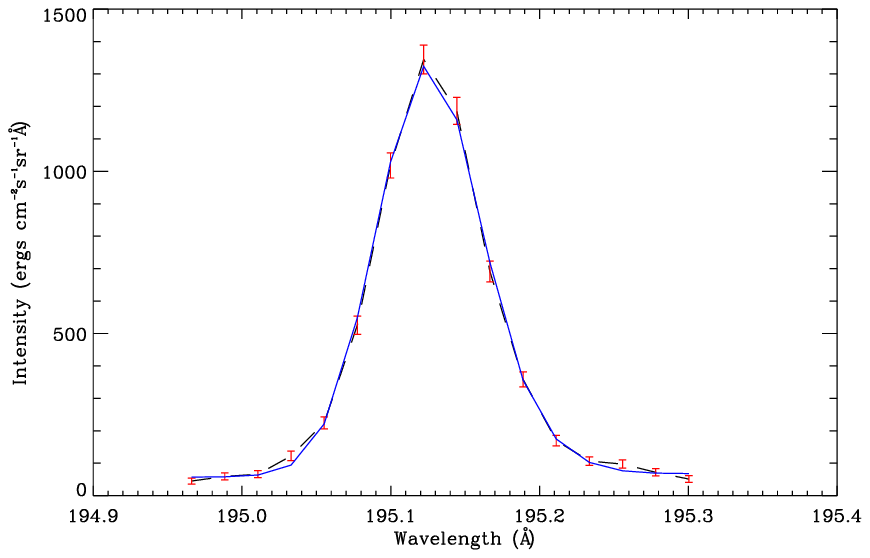}
}
\mbox{
\includegraphics[width=6.0cm]{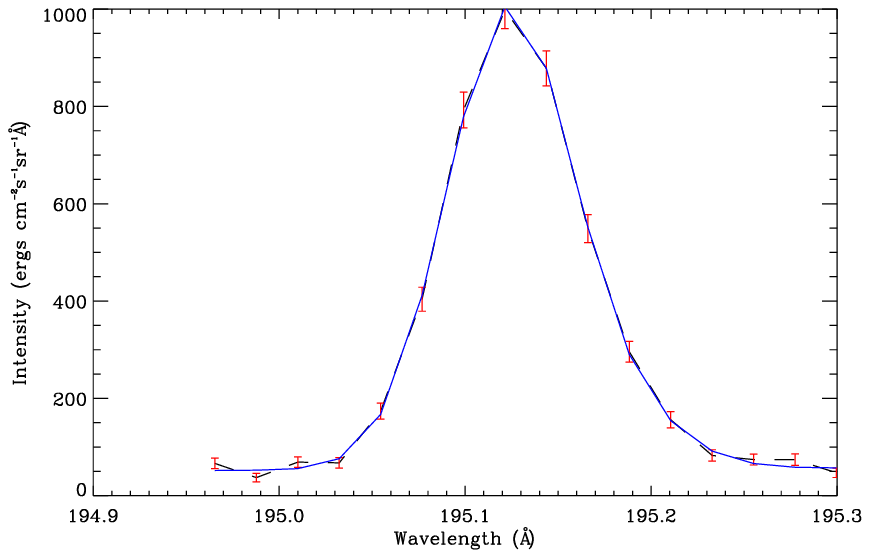}
\includegraphics[width=6.0cm]{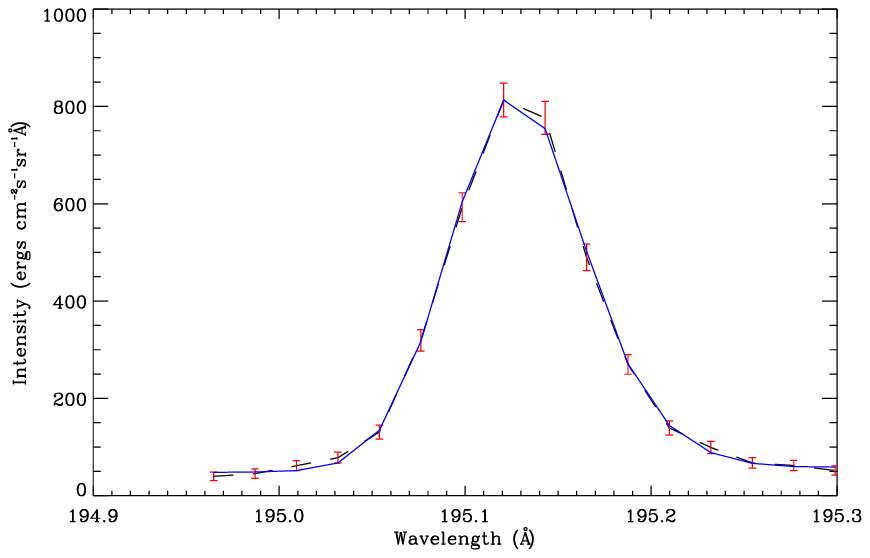}
\includegraphics[width=6.0cm]{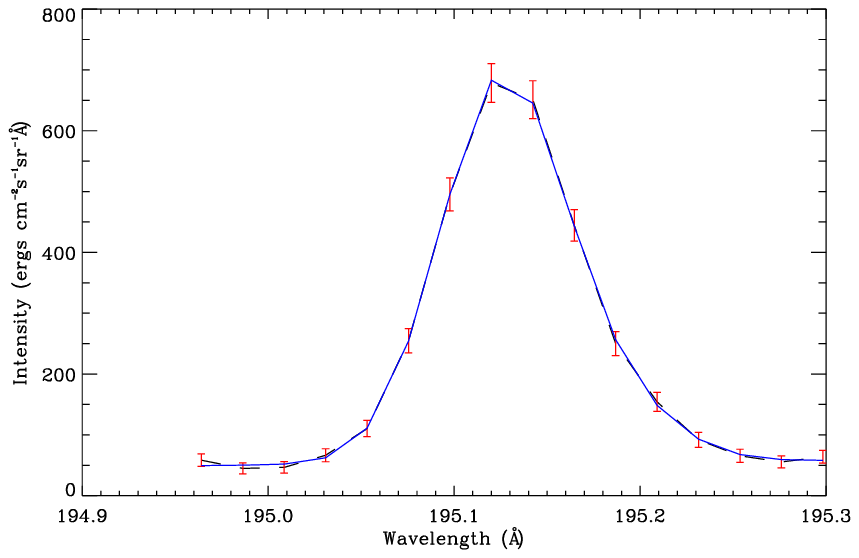}
}
\mbox{
\includegraphics[width=6.0cm]{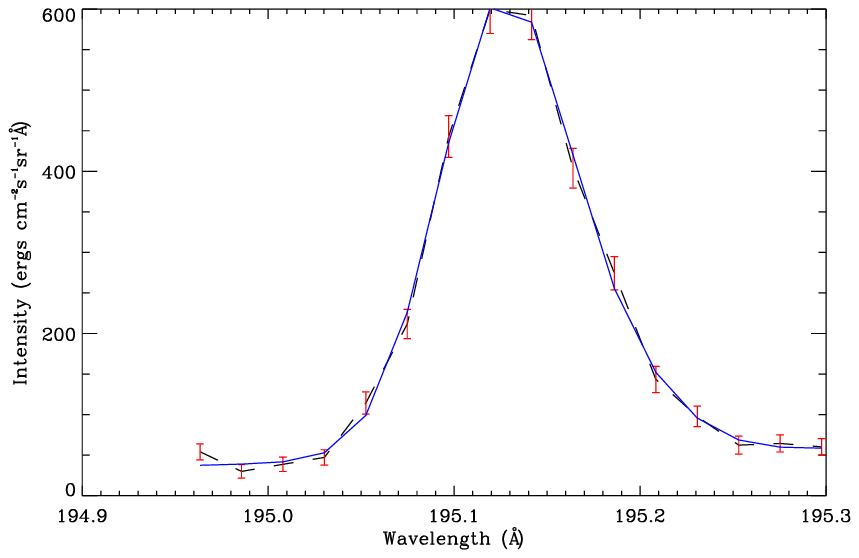}
\includegraphics[width=6.0cm]{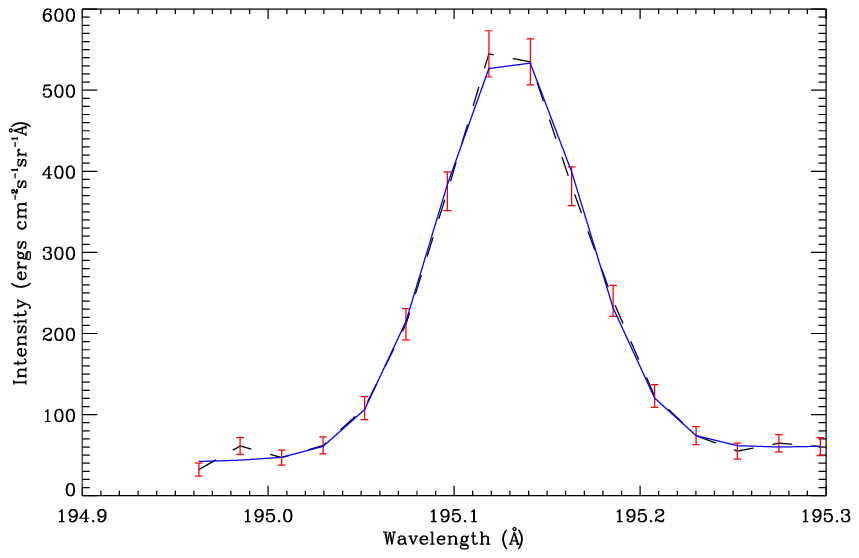}
\includegraphics[width=6.0cm]{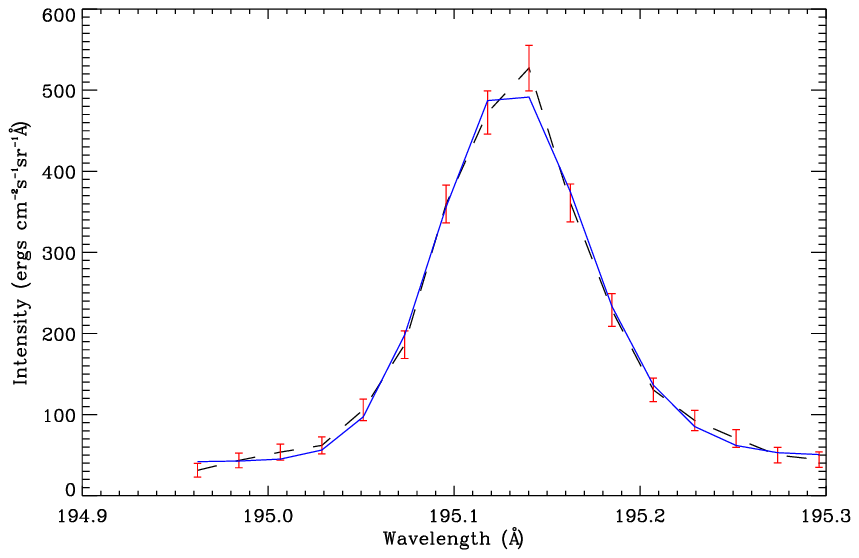}
}
\mbox{
\includegraphics[width=6.0cm]{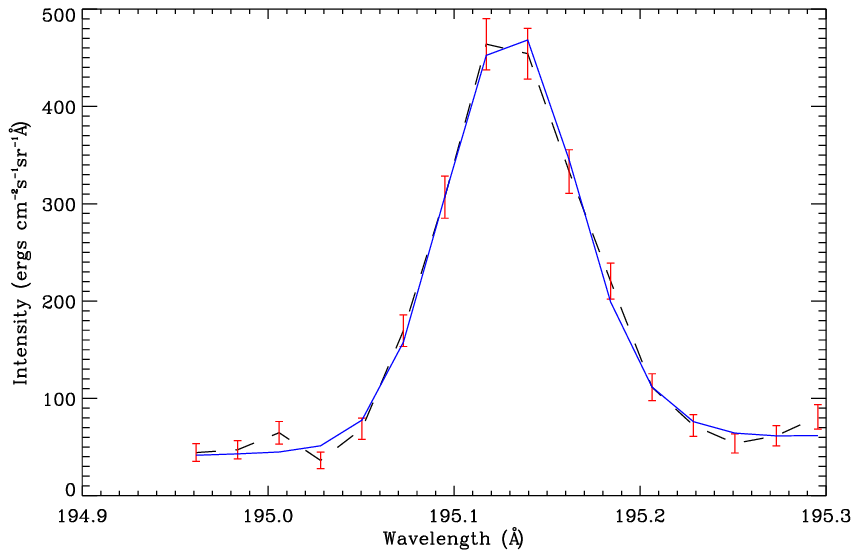}
\includegraphics[width=6.0cm]{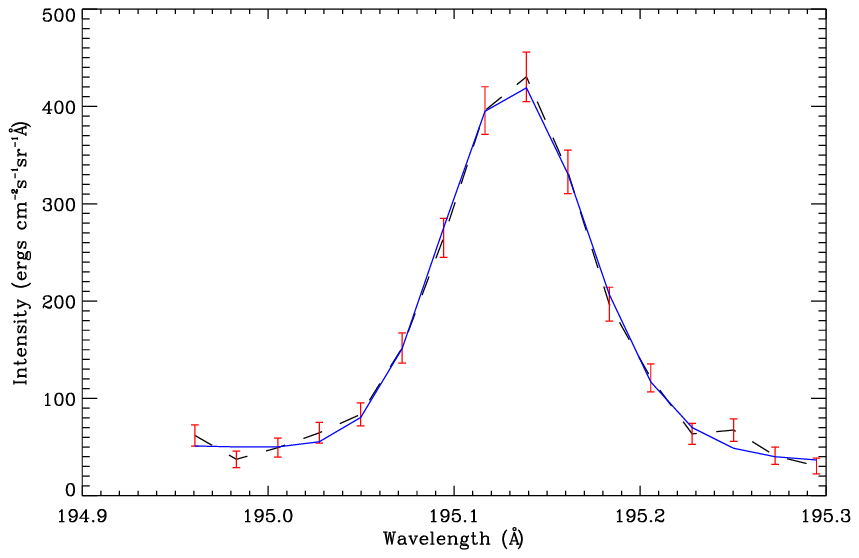}
}
\caption{Gaussian fit to the spectra derived on the chosen path along the jet. The white slit on the jet (cf., Fig.~1) is divided in $14$ super pixels each of the size of $2 \times 6$ pixel square. Line profile is derived  from each super pixel (dotted-blue), and corressponding estimation of plasma parameters are made by fitting the Double Gaussian (solid-blue).}
\label{fig304}
\end{figure*}

\section{Hinode/EIS Observations of a Polar Coronal Jet}
\label{obs}
\subsection{Spectroscopic data and observational analyses}
A polar jet is observed using $2"$ slit scan by EUV Imaging Spectrometer (Culhane et al. 2006) onboard Hinode on 22 April, 2009. It is an established fact that the $\mathbf{40"}$ and $\mathbf{266"}$ slots observe the temporal image data, while $\mathbf{1"}$ and $\mathbf{2"}$ slits are appropriate for the spectroscopic observations of the solar corona and Transition Region (TR). The EIS observes high-resolution spectra in two wavelength intervals, i.e., $170-211$ and $246-292~\mathrm{\AA}$ using its short-wavelength (SW) and long-wavelength (LW) CCDs respectively. 
The observed data contain spectral line profiles of Fe XIII $202.04~\mathrm{\AA}$, Fe XI $188.23~\mathrm{\AA}$, Fe XV $284.16~\mathrm{\AA}$, Fe XII $195.12~\mathrm{\AA}$, Fe X $184.54~\mathrm{\AA}$, Si VII $275.35~\mathrm{\AA}$, O V $192.9~\mathrm{\AA}$ and He II $256.32~\mathrm{\AA}$. The scanning started at 05:31:43 UT and ended at 06:33:29 UT. The exposure time on each scanning step was $31~\mathrm{s}$. The polar jet and related plasma column that was moving off the limb is scanned fully in a single spatio-temporal step at 06:29 UT because it went straight into the corona. This provides us an opportunity to understand the wave activity along the jet that already reached up to a certain height in the polar corona. The $2"$ slit started scanning steps over the polar coronal hole with (Xcen, Ycen) $\sim$ ($\mathbf{30"}$, $\mathbf{939"}$). The observation window on the CCDs is $400$ pixels high along the slit with a width of $40$ pixels in the horizontal direction. The Y-direction covers the solar atmosphere from $740"$ to $1139"$ (400 pixels with 1$"$/pxiel). 
The direction of dispersion has the spectral resolution of $0.02~\mathrm{\AA}$. 

We apply standard EIS data-reduction procedures and calibration files to the data acquired at the EUV-telescope which is the raw (zeroth-level) data. The subroutines can be found in the sswidl software tree\footnote{http://www.darts.isas.jaxa.jp/pub/solar/ssw/hinode/eis/}. These standard subroutines correct for dark-current subtraction, cosmic-ray removal, flat-field correction, hot pixels, warm pixels and bad/missing pixels. The data are saved in the level-1 data file, while associated errors are saved in the error file. We choose the strong line Fe XII $195.12~\mathrm{\AA}$ to examine the spatial variations of the intensity, Doppler velocity, and FWHM along the jet from its base in order to get the clues of transversal waves. We co-align the Fe XII $195.12~\mathrm{\AA}$ map w.r.t. the long-wavelength CCD observations of He II $256.32~\mathrm{\AA}$ by considering it as a reference image and by estimating the offset. The orbital and slit-tilt are also corrected for the data using the standard method described in the EIS software notes. We perform the double Gaussian fitting for the removal of the weak blend of Fe XII $195.18~\mathrm{\AA}$ line that affects the line profile of Fe XII $195.12~\mathrm{\AA}$. The fitting function is a Gaussian for line profile (see gauss\_.pro in SolarSoft), while straight-line for background continuum (see line\_.pro in SolarSoft). {\bf The fitting of the observed spectral line profile gives peak intensity, centroid (measure of flows), and Gaussian width (measure of thermal and non-thermal motions).} We apply the procedure described by Young et al. (2009) in this context, which is also available in the EIS Software Note 17. We constrain Fe XII 195.18 \AA line to have the same width as of Fe XII $195.12~\mathrm{\AA}$ line. We search for the blend line Fe XII $195.18~\mathrm{\AA}$ within the range of $+0.06~\mathrm{\AA}$ w.r.t. the centeroid of the main line Fe XII $195.12~\mathrm{\AA}$. We assume and fix that the contribution of the blended line will be searched within maximum $28~\%$ limit compared to the peak intensity of the main Fe XII $195.12~\mathrm{\AA}$ line. We perform the fitting over binned data of $2$ pixel $\times$ $6$ pixel to increase the signal-to-noise ratio and to obtain the resonable fitting and estimated parameters (cf., Fig.~1 and Fig.~2). The example of the Fe XII $195.12~\mathrm{\AA}$ fitted spectra at $14$ points along the chosen slit along jet are shown in Fig.~2. These fitted profiles of Fe XII $195.12~\mathrm{\AA}$ are free from any contribution of the weak blend of Fe XII $195.18~\mathrm{\AA}$. 
The $14$ spatial points along the jet corresspond to various heights for which the spectral parameters are derived (Fig.~3).

It should be noted that the wavelength calibration and estimation of the reference wavelength of Fe XII $195.12~\mathrm{\AA}$ line is performed using the limb method (Peter \& Judge 1999). The box is chosen at the north pole limb from where the integrated spectra is derived (cf., Fig.1, intensity map). It is likely that in and outward motions near the limb will almost cancel to each other. Therefore, the centroid of the line profile will represent the rest wavelength of particular emission line. The Gaussian fitting gives the estimate of the reference wavelength as $195.128~\mathrm{\AA}$. This reference wavelength is used in deriving the Doppler velocities in the region of our interest (cf., Figs.~1 and 3).

\subsection{Observational results}
 
Fig.~1 displays the intensity (left), Doppler-velocity (middle), FWHM (right) maps of the polar coronal hole where the straight coronal jet has already moved off the limb. The maps are related with spectral measurements over $2$ pixel $\times$ $6$ pixel binned data with good signal-to-noise ratio. The jet is visible in the left-most part of the intensity map, and it has a typical inverted Y-shape structure {\bf (cf., within the box in the intensity map)}. At the same place, in the Doppler velocity map, we see the red-shifted base of the jet while the plasma moves outward (blue-shift) in the higher plasma column of the jet {\bf (cf., within the box in the Doppler velocity map)}. The blue and red-shift measured in the off-limb jet corresspond to the line-of-sight component of velocity variations. It seems that the jet is slightly tilted towards us. Therefore, at a certain height, the jet plasma which is outflowing towards us provides a blue-shift signature, while the downflowing plasma towards the Sun within the jet gives the red-shift.  In the FWHM map, we find the increased contribution of the line-width in the jet plasma column {\bf (cf., within the box in the line-width map)}.
It should also be noted that the jet is not visible in Si VII 275.35 line (Log T$_{f}\approx5.0$), thereby confirming that it is not formed by cool transition region plasma. The jet is also not evident at higher temperatures. It is found that mostly its plasma is maintained at the inner coronal temperature of $1.0~\mathrm{MK}$. Therefore, the jet is best observed in Fe XII $195.12~\mathrm{\AA}$ line. We do not conjecture that it is made up of a single temperature plasma. In the given observational base-line its plasma temperature can on average be around the formation temperature of Fe XII line, i.e., Log $\mathrm{T}_{f}\approx6.1$. 

\begin{figure*}
\centering
\mbox{
\includegraphics[width=8.5cm]{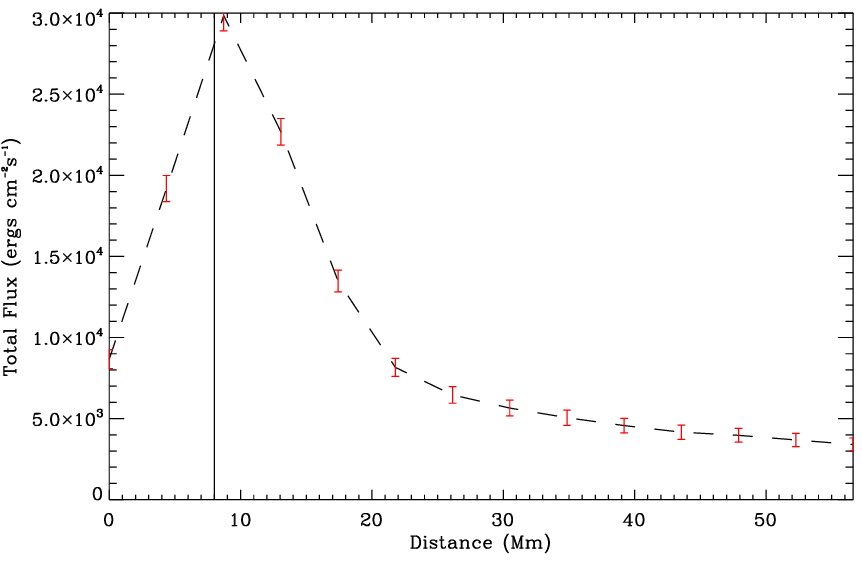}
\includegraphics[width=8.5cm]{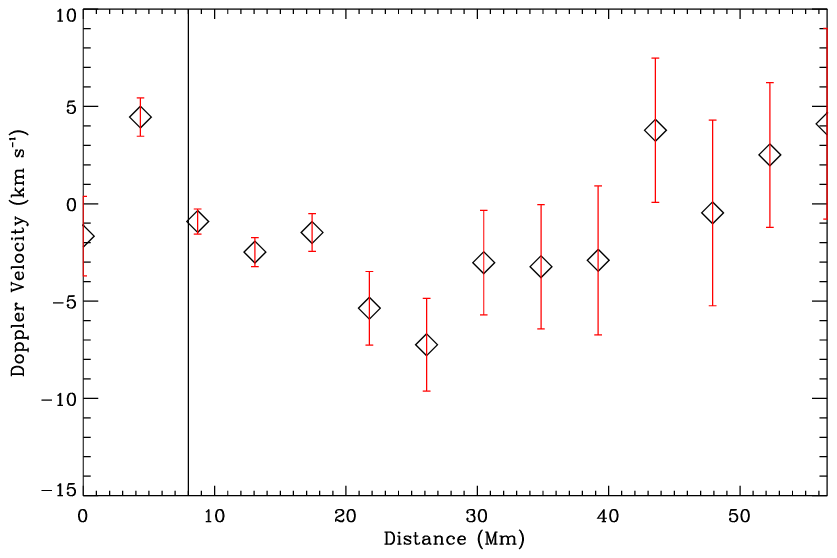}}
\includegraphics[width=8.5cm]{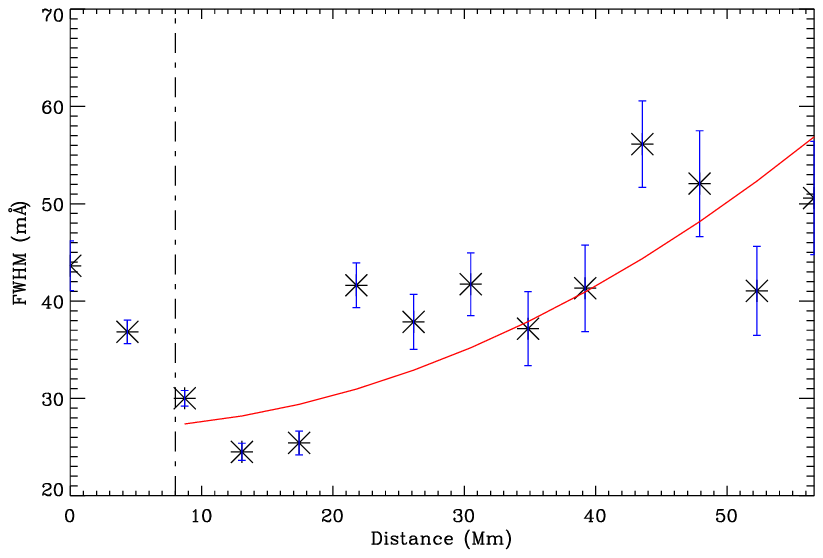}


\caption{The variation of intensity, Doppler velocity, and FWHM of Fe XII 195.12 \AA\ line along the path chosen on the polar jet from its base outward. Red-line shows third order polynomial fit to the FWHM data.}
\label{fig304}
\end{figure*}

We choose a straight slit from the base of the jet in the outward direction along the jet plasma column (cf., Fig.~1). We measure the variation of intensity (top-left panel), Doppler velocity (top-right panel), and FWHM (bottom panel) of Fe XII $195.12~\mathrm{\AA}$ line along this chosen path (cf., Fig.~3).
The first position at the bottom of the slit is the base of the jet which is referenced as 'zero Mm'. The rest of the slit represents the progressive height along the jet scaled in Mm. The total height of the slit along the jet is 
$56~\mathrm{Mm}$ (cf., Fig.~3). It should be noted that we have considered the estimations of these spectral parameters along the jet up to only those heights $(\approx 55~\mathrm{Mm})$ where spectral fitting is resonable with less fitting errors (cf., Fig.~2). 
It is clear from these panels that intensity is continously enhanced upto $10~\mathrm{Mm}$ and decreases thereafter Doppler velocity changes from its maximum value of red-shift about $+5~\mathrm{km\,s}^{-1}$ from the base of the jet to the blue-shift (outflows) at a height between $5-10~\mathrm{Mm}$. We find the variation of FWHM along the same path. It is almost constant without its much variation upto $10-15~\mathrm{Mm}$ (cf., Fig.~3, bottom panel). Above this height, the FWHM increases from 30 m\AA\ to 55 \AA\ with a significant positive gradient (cf., red fitted line in the bottom panel of Fig.~3). Red-line shows third order polynomial fit on the FWHM data.
\subsection{Physical interpretations of the observational results}
%
{\bf The small-scale and low-lying bright-point loops form the wide base of the jet. They reconnect with pre-existing open field lines and propell jets' plasma.} This scenario is in accordance with the model given by Yokoyama \& Shibata (1995) about the coronal jet formation. The red-shift is maximum ($+5~\mathrm{km\,s}^{-1}$) at $5~\mathrm{Mm}$ height, which means the plasma is downflowing at this location. Above this height between $5-10~\mathrm{Mm}$, this red-shift is transformed into blue-shift. This vertical region acts as a reconnection region where plasma flows in both directions: the downflows in small-scale loops forming the base of the jet and upflows along open field lines of the jet forming its plasma motion higher into the corona. This region where downflows (red-shift) convert into outflows (blue-shift) and intensity (thus density) tends toward its maximization, indicates that reconnection already drove the energy release below of this location near the base of the jet. The reconnection propells the plasma outwards along the open field lines of the jet. The maximum observed blue-shift is $7.5~\mathrm{km\,s}^{-1}$ along the jet. However, it should be noted that the jet is projected off the limb at the north-pole. The plasma motions along the jet may have minimum contribution to the line-of-sight (LOS) Doppler shift. Therefore, the estimated Doppler velocity should be considered as an apparent lower-bound velocity, which is not as an actual speed of the jet. This simply provides the signature of the outflowing plasma along the jet.
The FWHM also shows the increasing trend beyond $10-15~\mathrm{Mm}$ height. It should be noted that the FWHM value exactly at the base of the jet at the north polar limb is somewhat higher, i.e., $40~\mathrm{m\AA}$. We can discard it due to the limb effects. The contribution of stray light from continuum near the limb may affect the line profile and its width. Second point of estimated FWHM still lies higher around $36~\mathrm{m\AA}$ at $5~\mathrm{Mm}$ where the highest red-shift (downflow) is observed. At $5.0~\mathrm{Mm}$ height, the energy release is likely to be present as downflow is maximum at this location. The energy release may cause the line-width broadening of the spectral line (Somov 1994). 
Therefore, the spatial points that lie in the lower segment of the jet closer to the 
energy release site in the spatial range $5-15~\mathrm{Mm}$ from Jet's footpoint, possess comparatively higher values of FWHM.
As we move away from energy release site at higher heights, the jet's temperature is uniform near the formation temperature of optically thin Fe XII $195.12~\mathrm{\AA}$ line. Therefore, the contribution to the  FWHM increment may be mainly due to the increment in non-thermal motions.
 
The reconnection generates the outflowing plasma along jet's open magnetic field lines, as well as the downflowing plasma along the small-scale loops forming its base. This region ($0-15~\mathrm{Mm}$), therefore, may be the most likely place in the lower segment  of the jet around which the reconnection and related physical processes occurred. Beyond this region, the FWHM of the observed spectral line increases significantly due to non-thermal motions as evident in the observational base-line. This FWHM broadening may be associated with the evolution and growth of the non-linear and impulsively generated Alfv\'en waves propagating along the open field lines of the jet away from the reconnection site higher into the corona.
In the theory as well as observations, Alfv\'en waves are found to be excited in the large-scale polar coronal holes (Banerjee et al. 1998, Chmielewski et al. 2013).

The FWHM also shows varying trend in and around the observed jet. However, these trends are not so systematic and closely correlated with intensity and Doppler velocity variations with height as we observed within the jet.
We do not present these plots. However, we have examined the region outside the jet's plasma column and its inverted Y-shape base in its western side. The intensity decreases rapidly as we move outside the limb, which is an obvious trend. However, there is no trend in the Doppler velocity. Doppler velocity only shows weak downflows. FWHM also does not show any trend in its variation with height.
In conclusion, present observations demonstrate the signature of impulsive  Alfv\'en waves along the jet above the energy release site near its footpoint. In the next section, we describe our model of impulsive Alfv\'en waves in the 2-D vertical current-sheet to match with some observed properties of these wave modes.

\section{Numerical model of impulsive Alfv\'en waves in 2-D vertical Harris current-sheet}
\label{num}

Keeping in mind the impulsive generation of Alfv\'en waves due to energy release within the observed jet, we model the propagation of a transverse pulse in a 2-D vertical Harris current-sheet. Above the reconnection point, perturbations may evolve in the form of Alfv\'en waves propagating at various higher heights as well as the driven coupled plasma flows. We consider the height of $5~\mathrm{Mm}$ in our model to launch the reconnection generated velocity pulse in a 2-D vertical current-sheet.
\subsection{Governing equations}
In our numerical model of a Harris current-sheet, embedded in a gravitationally stratified solar atmosphere, the plasma dynamics is described by the two and half-dimensional (2.5-D) time-dependent ideal MHD equations (cf., Priest, 1982; Chung, 2002):
\begin{equation}\label{eq1}
\frac{\mathrm{D}\varrho}{\mathrm{D}t} = -\varrho \nabla\cdot \bm{v},
\end{equation}
\begin{equation}\label{eq2}
\varrho \frac{\mathrm{D}\bm{v}}{\mathrm{D}t} = -\nabla
p+\bm{j}\times\bm{B} + \varrho \bm{g},
\end{equation}
\begin{equation}\label{eq3}
\frac{\mathrm{D}\bm{B}}{\mathrm{D}t} = (\bm{B} \cdot \nabla)\bm{v},
\end{equation}
\begin{equation}\label{eq4}
\frac{\mathrm{D}e}{\mathrm{D}t} = -\gamma e \nabla \cdot \bm{v},
\end{equation}
\begin{equation}\label{eq5}
\nabla\cdot\bm{B}=0.
\end{equation}
Here $\mathrm{D} / \mathrm{D}t \equiv \partial / \partial t + \bm{v} \cdot \nabla$ is the total time derivative, $\varrho$ is the mass density, $\bm{v}$ is the flow velocity, $\bm{B}$ is the magnetic field, and $\bm{g} = [0,-g_{\sun},0]$ is the gravitational acceleration, that is oriented in the negative $y$-direction, with $g_{\sun} = 274~\mathrm{m \cdot s^{-2}}$. Our model does not include the radiative and thermal conductive losses. This model may not fully describe the reconnection generated heating and related additional plasma evolution within the jet as we do not invoke the non-classical thermodynamical terms (e.g., radiative and thermal losses) in the governing energy equation. In the present paper, we do aim to understand only the non-linear wave dynamics and associated vertical flows along the jet.

The current density $\bm{j}$ in Eq. (\ref{eq2}) is expressed as
\begin{equation}\label{eq6}
\bm{j} = \frac{1}{\mu_0}(\nabla \times \bm{B}),
\end{equation}
where $\mu_0 = 1.26 \times 10^{-6}~\mathrm{H\cdot m}^{-1}$ is the magnetic permeability of free space. 
 
The 
internal energy 
density, $e$, in Eq. (\ref{eq4}) is given by
\begin{equation}\label{eq7}
e = \frac{p}{\gamma - 1},
\end{equation}
with the adiabatic coefficient which we set and hold fixed as $\gamma = 5/3$.
\subsection{Gravitationally stratified Harris current-sheet}
For a static ($\bm{v} = \bm{0}$) equilibrium, the Lorentz and gravity forces must be balanced by the pressure gradient in the entire physical domain
\begin{equation}\label{eq8}
-\nabla p+\bm{j}\times\bm{B} + \varrho \bm{g} = \bm{0}.
\end{equation}
The solenoidal condition, $\nabla\cdot\bm{B}=0$, is identically satisfied by the
magnetic flux function, $\bm{A}$,
\begin{equation}\label{eq9}
\bm{B} = \nabla \times \bm{A}.
\end{equation}
For calculating the magnetic field in the vertically oriented Harris current-sheet, we use the magnetic flux function $\bm{A} = [0,0,A_z]$ as (Galsgaard \& Roussev, 2002; Jel\'\i nek et al., 2012)
\begin{equation}\label{eq10}
A_z = -B_{\mathrm{0}} w_\mathrm{cs} \ln\left\{\left|\cosh\left(\frac{x}{w_\mathrm{cs}}\right)\right|\right\}\exp \left(-\frac{y}{\lambda}\right).
\end{equation}
Here the coefficient $\lambda$ denotes the magnetic scale-height. The symbol $B_{\mathrm{0}}$ is used for the external magnetic field and $w_\mathrm{cs}$ is the half-width of the current-sheet. We set and hold fixed $w_\mathrm{cs} = 1.0~\mathrm{Mm}$. 
Substituting Eq.~(10) into Eq.~(9) we obtain the equations for the magnetic field in the $x$-$y$ plane as (Galsgaard \& Roussev, 2002; Jel\'\i nek et al., 2012)
\begin{equation}\label{eq15}
B_x(x,y) = B_{\mathrm{0}} \frac{w_\mathrm{cs}}{\lambda} \ln \left[\cosh \left(\frac{x}{w_\mathrm{cs}}\right)\right] \exp \left(-\frac{y}{\lambda}\right),
\end{equation}
\begin{equation}\label{eq16}
B_y(x,y) = B_{\mathrm{0}} \tanh \left(\frac{x}{w_\mathrm{cs}}\right)\exp \left(-\frac{y}{\lambda}\right).
\end{equation}

\begin{figure}
\includegraphics[scale = 0.44]{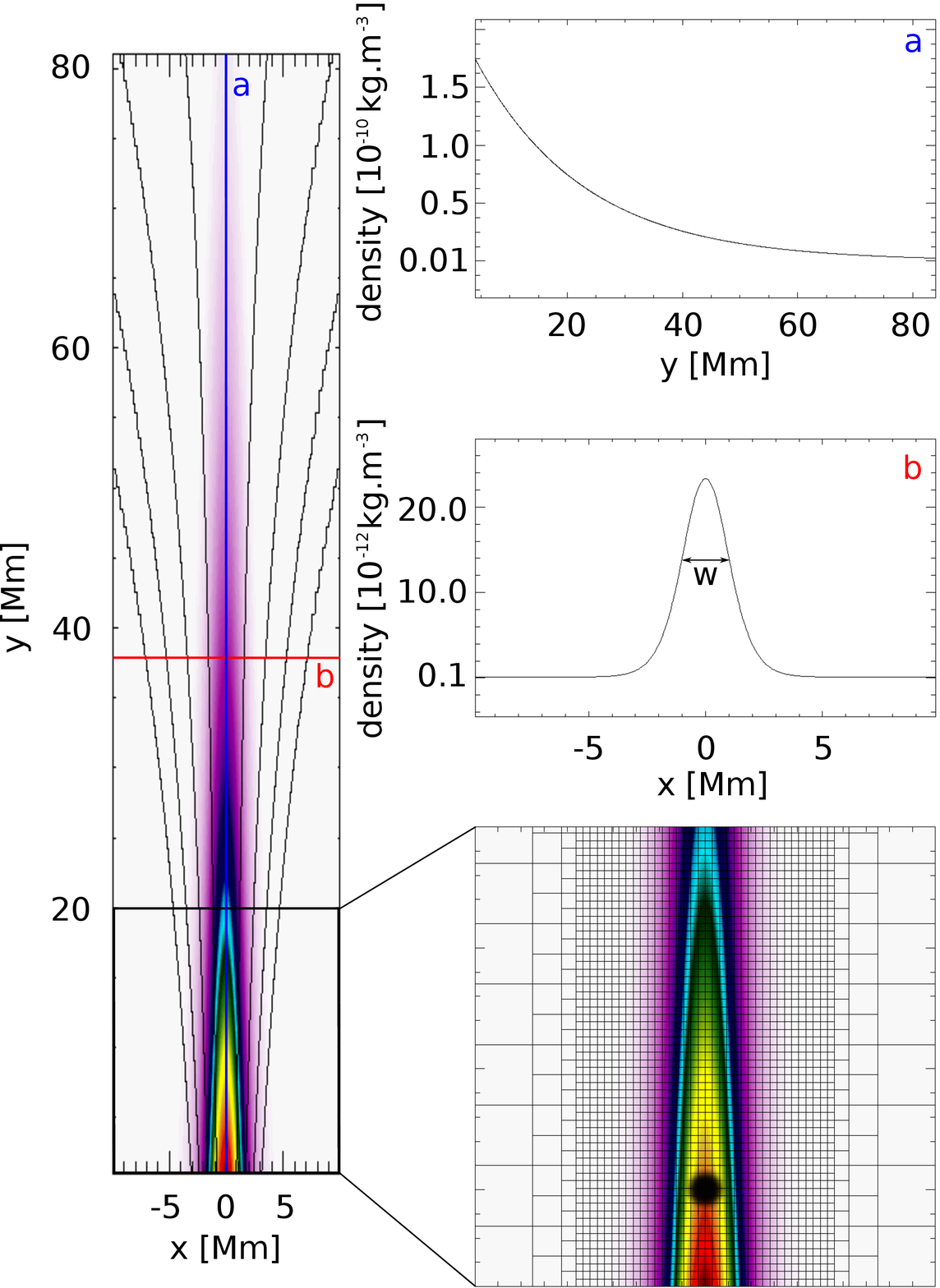}
\caption{The mass density distribution in the Harris current-sheet, where the black contours represent the structure of magnetic field. 
On the right-hand side of the figure both (a) vertical (for $x = 0~\mathrm{Mm}$) and (b) horizontal 
(for $y = 37.65~\mathrm{Mm}$, which is magnetic scale height)}
slices in mass density are shown, as well as the details of the part of the simulation region illustrating the computational grid, with Adaptive Mesh Refinement (AMR). At the altitude $y=5~\mathrm{Mm}$ the perturbation point, $L_\mathrm{P}$, is marked by full black circle.
\label{fig304}
\end{figure}
 
Equation (8), while written in terms of its components, attains the following form:
\begin{equation}\label{eq17}
\frac{\partial p(x,y)}{\partial x} + j_z B_y(x,y) = 0,
\end{equation}
\begin{equation}\label{eq18}
\frac{\partial p(x,y)}{\partial y} - j_z B_x(x,y) + \varrho g(x,y) = 0.
\end{equation}
Here $j_z$ is the only non-zero component of the electric current density $\bm{j}$ (see Eq. (\ref{eq6})), given by $j_z = \frac{1}{\mu_0}(\nabla \times \bm{B})_z$.
The condition of integrability of the above equations leads to
\begin{equation}\label{eq19}
\mu_0 g \frac{\partial \varrho(x,y)}{\partial x} = \nabla \cdot ( \mu_0 j_z \bm{B}),
\end{equation}
from which we can derive the formulae for the distribution of the mass density (cf., Galsgaard \& Roussev, 2002; Jel\'\i nek et al., 2012)
\begin{eqnarray}\label{eq20}
\varrho(x,y) &=& \Bigg\{\frac{B_\mathrm{0}^2}{\mu_0 g \lambda}\left\{1+\ln\left[\cosh\left(\frac{x}{w_\mathrm{cs}}\right)\right]\right\}\mathrm{sech}^2\left(\frac{x}{w_\mathrm{cs}}\right) + \nonumber \\
&+&\varrho_0 \Bigg\} \cdot \exp{\left(-2\frac{y}{\lambda}\right)},
\end{eqnarray}
and a gas pressure
\begin{eqnarray}\label{eq21}
p(x,y) &=& \Bigg\{\frac{B_\mathrm{0}^2}{2 \mu_0} \mathrm{sech}^2\left(\frac{x}{w_\mathrm{cs}}\right) + \frac{B_\mathrm{0}^2 w_\mathrm{cs}^2}{2 \mu_0 \lambda^2} \ln^2\left[\cosh\left(\frac{x}{w_\mathrm{cs}}\right)\right] +
\nonumber \\
&+&\frac{\varrho_0 g \lambda}{2}\Bigg\} \cdot \exp{\left(-2\frac{y}{\lambda}\right)} + p_0.
\end{eqnarray}
Here $\varrho_0$ and $p_0$ are integration constants. The corresponding plasma temperature is assumed to be $T = 1.25 \times 10^6~\mathrm{K}$.

\subsection{Initial perturbation}
At the start of the numerical simulation ($t = 0~\mathrm{s}$), the equilibrium is perturbed by the Gaussian pulse in the $z$-component of velocity and has the following form (e.g. Nakariakov et al. 2004, 2005):
\begin{equation}
V_z(x,y, t = 0) = - A_0 \cdot \frac{x}{\lambda} \cdot \exp{\left[-\frac{x^2 + (y-L_{\mathrm{P}})^2}{\lambda^2}\right]},
\end{equation}
where $A_0$ is the amplitude of the initial pulse, and $\lambda = 4~\mathrm{Mm}$ is its width. This pulse preferentially triggers Alfv\'en waves. The perturbation point, $L_\mathrm{P}$, is located on the axis of the Harris current-sheet, at a distance of $5~\mathrm{Mm}$ from the bottom boundary of the simulation region (see full black circle in Fig.~4). We set the detection points, $L_\mathrm{D}$, on the current-sheet axis, at the distance between the perturbation and detection points $\Delta \equiv |L_\mathrm{D} - L_\mathrm{P}| = 10, 15, 20, 25, 30, 35, 40, 45, 50$, and $55~\mathrm{Mm}$.

Note that in the 2.5-D model, the Alfv\'en waves decouple from magnetoacoustic waves. They can be described solely by $V_z(x,y,t)$. 
As a result, the initial pulse of Eq.~(18) triggers Alfv\'en waves that are approximately described in the linear case by the wave equation
\begin{equation}
\frac{\partial^2 V_z(x,y,t)}{\partial t^2} = c_\mathrm{A}^2(x,y) \frac{\partial^2 V_z(x,y,t)}{\partial s^2},
\end{equation}
where $s$ is the coordinate along the magnetic line and the Alfv\'en speed, $c_\mathrm{A}$, is defined as
\begin{equation}
c_\mathrm{A}(x,y) = \sqrt{\frac{B_x^2(x,y)+B_y^2(x,y)}{\mu_0 \varrho(x,y)}}.
\end{equation}
Here $\varrho(x,y)$ is given by Eq. (16).

\subsection{Numerical solutions}

We solve the 2.5-D time-dependent, ideal MHD equations (\ref{eq1})-(\ref{eq4}) numerically, making use of the FLASH code (Fryxell et al. 2000; Lee \& Deane 2009; Lee 2013). It is now well tested, fully modular, parallel, multiphysics, open science, simulation code that implements second- and third-order unsplit Godunov solvers with various slope limiters and Riemann solvers as well as adaptive mesh refinement (AMR; see e.g. Chung 2002). The Godunov solver combines the corner transport upwind method for multi-dimensional integration and the constrained transport algorithm for preserving the divergence-free constraint on the magnetic field (Lee \& Deane 2009). We have used the minmod slope limiter and the Riemann solver (e.g. Murawski, 2002; Toro 2006). The main advantage of using AMR technique is to refine a numerical grid at steep spatial profiles while keeping a grid coarse at the places where fine spatial resolution is not essential. In our case, the AMR strategy is based on controlling the numerical errors in a gradient of mass density that leads to reduction of the numerical diffusion within the entire simulation region.

For our numerical simulations, we use a 2-D Eulerian box of its height $H = 80~\mathrm{Mm}$ and width $W = 20~\mathrm{Mm}$. The spatial resolution of the numerical grid is determined with the AMR method. We use the AMR grid with the minimum (maximum) level of the refinement blocks set to 3 (7). The whole simulation region is covered by $12884$ blocks. Since every block consists $8 \times 8$ numerical cells, this number of blocks corresponds to $824576$ numerical cells. 

Note that a spatial grid size has to be less than the typical width of the current-sheet along the $x$-direction and the typical wavelength of the Alfv\'en waves along the $y$-direction, respectively. We find $\min(\Delta x) = \min(\Delta y) = 0.04~\mathrm{Mm}$, which satisfies the above mentioned condition (semi-width of the current-sheet is $w_\mathrm{CS} = 1.0~\mathrm{Mm}$ and the estimated minimal wavelength is approximately $10.0~\mathrm{Mm}$). At all boundaries, we fix all plasma quantities to their equilibrium values using fixed-in-time boundary conditions, which lead only to negligibly small numerical reflections of incident wave signals.

\section{Results}

\subsection{Numerical results}

In a current-sheet, a velocity perturbation is generated most likely due to reconnection at a height of $5~\mathrm{Mm}$ (cf., Fig.~4), which evolves with height in the stratified solar atmosphere. The initial pulse excites Alfv\'en waves. These waves, which are associated with the perturbations in $V_{z}$, propagate along the open magnetic field lines to higher altitudes ($y$-direction): see the time evolution of $V_z$ at $t = 0, 10, 30,$ and $70~\mathrm{s}$ (Fig.~5). The initial pulse (left) spreads into upwardly and downwardly propagating waves which are very well seen at $t = 10~\mathrm{s}$ (second left). Since these waves follow magnetic field lines, which are diverging with height, the left and right profiles of the Alfv\'en waves move apart with $y$ (see the right most panel). 

\begin{figure*}
\centering
\mbox{
\includegraphics[scale = 0.4]{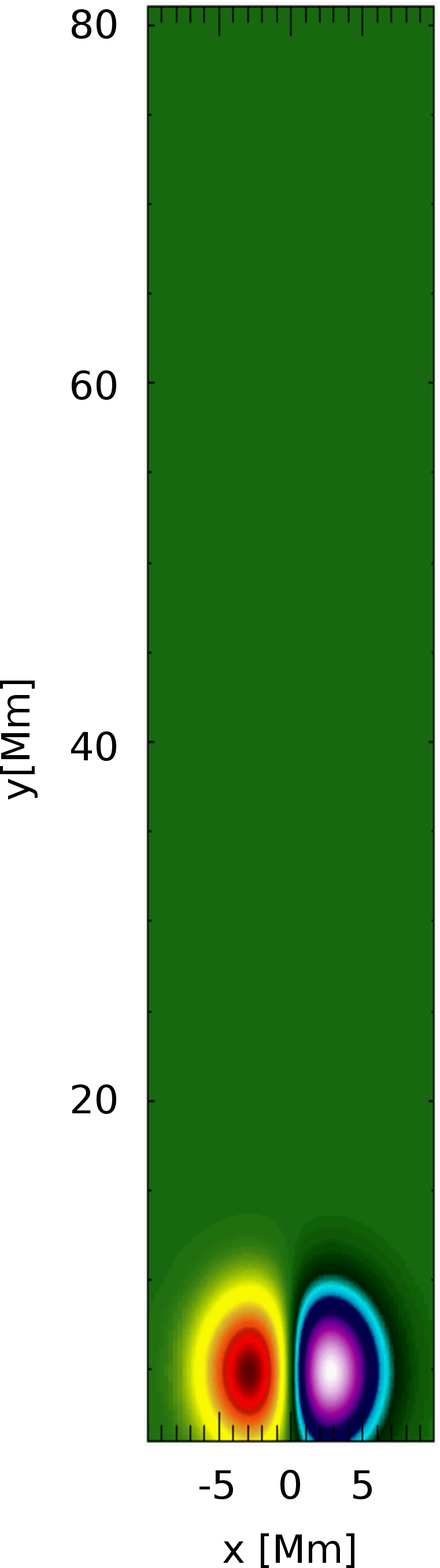}
\hspace{0.75cm}
\includegraphics[scale = 0.4]{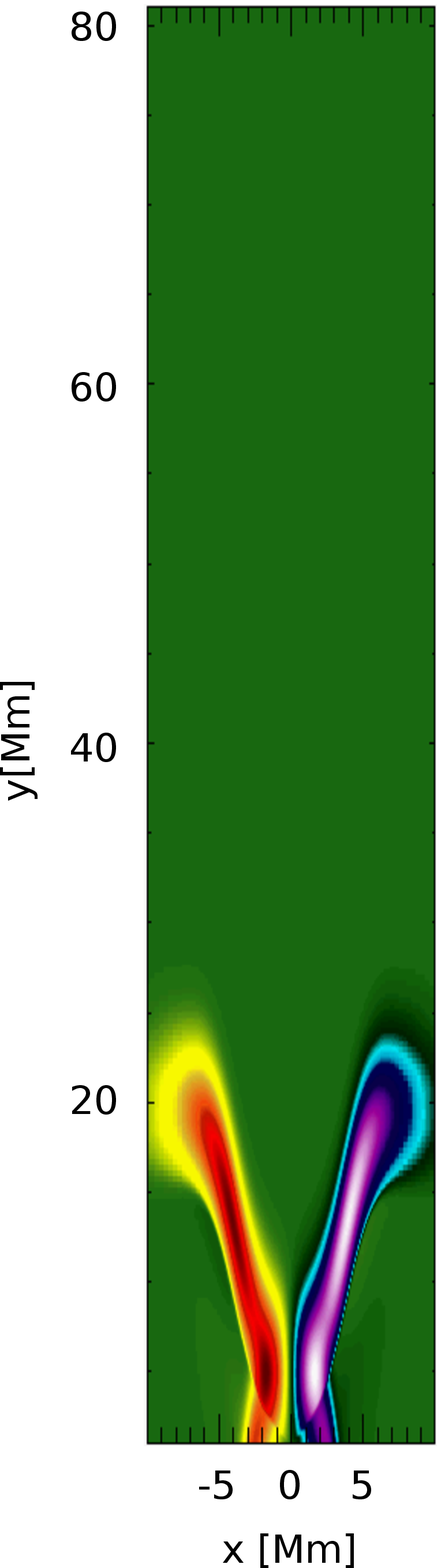}
\hspace{0.75cm}
\includegraphics[scale = 0.4]{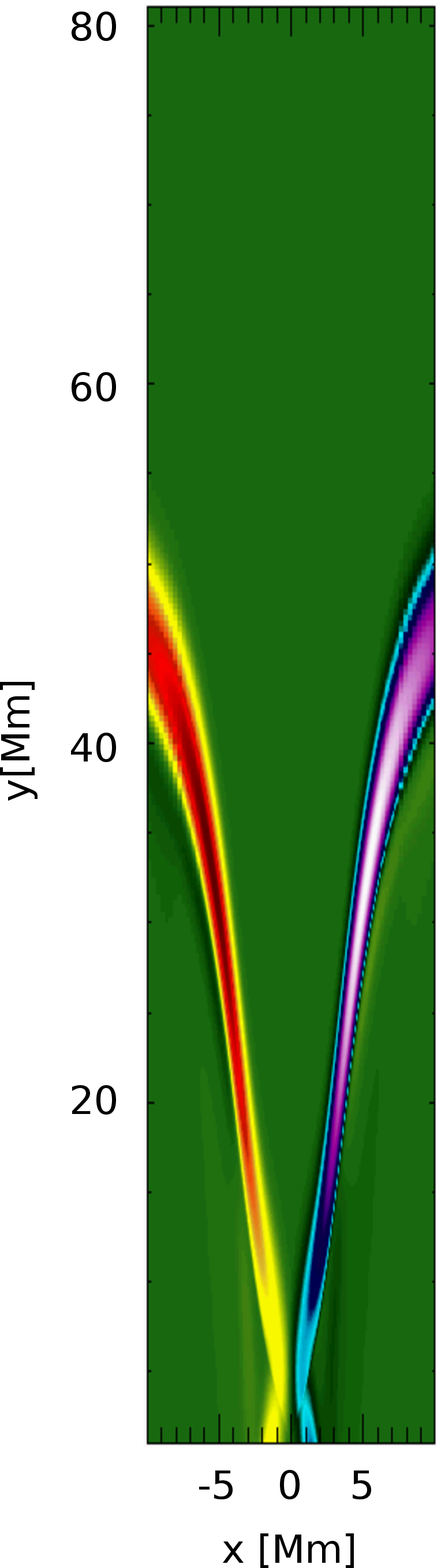}
\hspace{0.75cm}
\includegraphics[scale = 0.4]{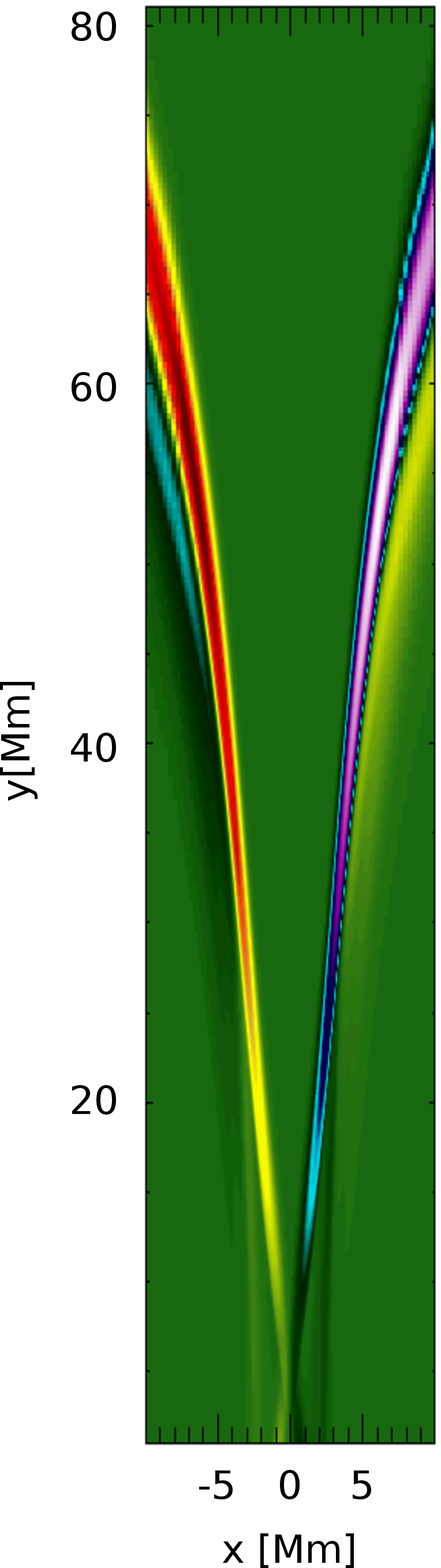}
}

\caption{The time evolution of velocity component $V_z$, corresponding to Alfv\'en waves, at $t = 0, 10, 30$, and $70~\mathrm{s}$ from left to right.}
\label{fig304}
\end{figure*}

The associated magnetoacoustic waves produce the mass density perturbations at various heights and also create the vertical plasma dynamics. Some instantaneous mass density, $V_{y}$ and $V_{z}$ velocity variations at $15, 35$, and $55~\mathrm{Mm}$ heights along the current-sheet above the solar surface are shown in Fig.~6. At each height, we have the temporal signatures of the perturbations of these parameters due to the evolution of waves. 

\begin{figure*}
\centering
\mbox{
\includegraphics[scale = 0.233]{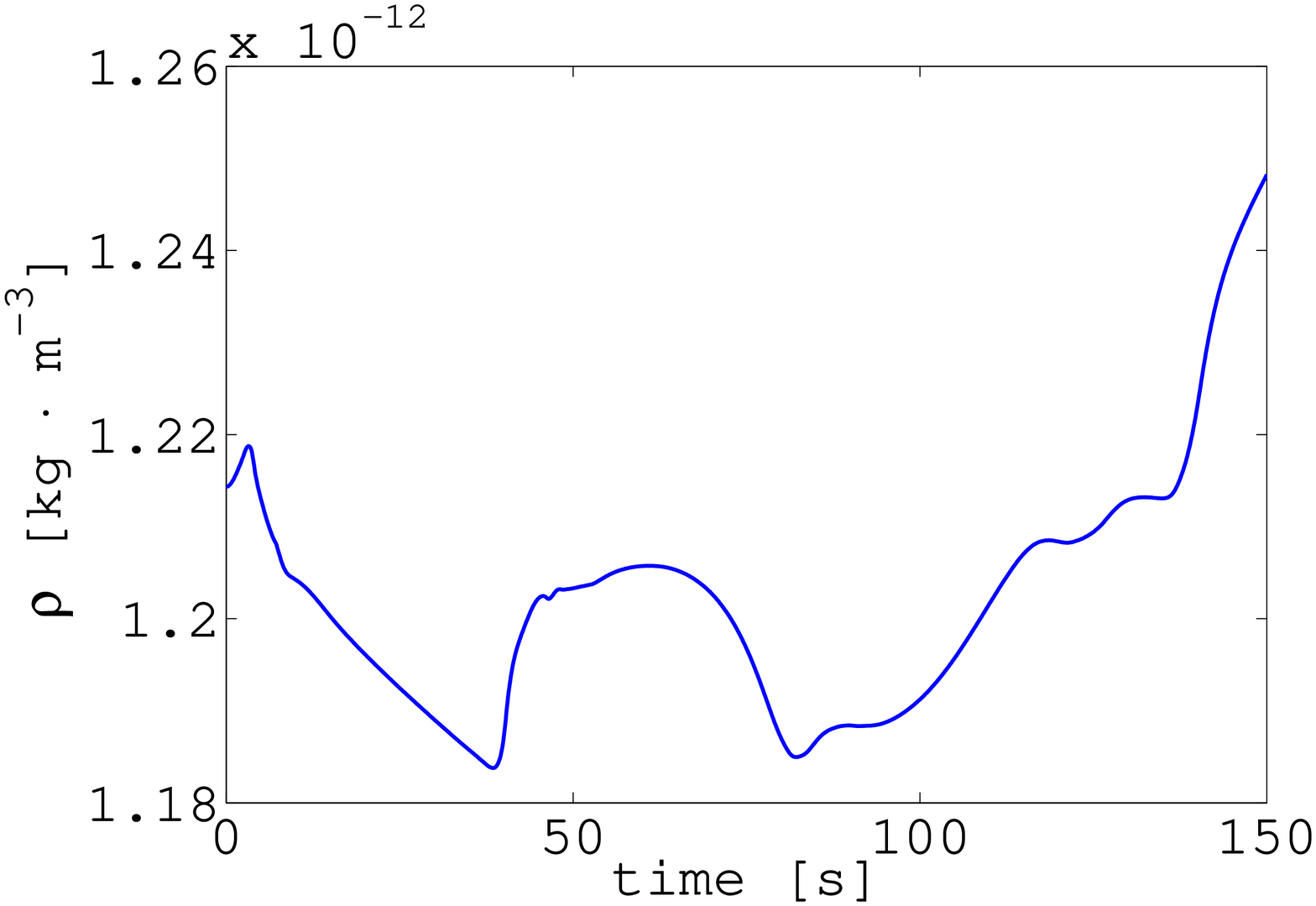}
\includegraphics[scale = 0.233]{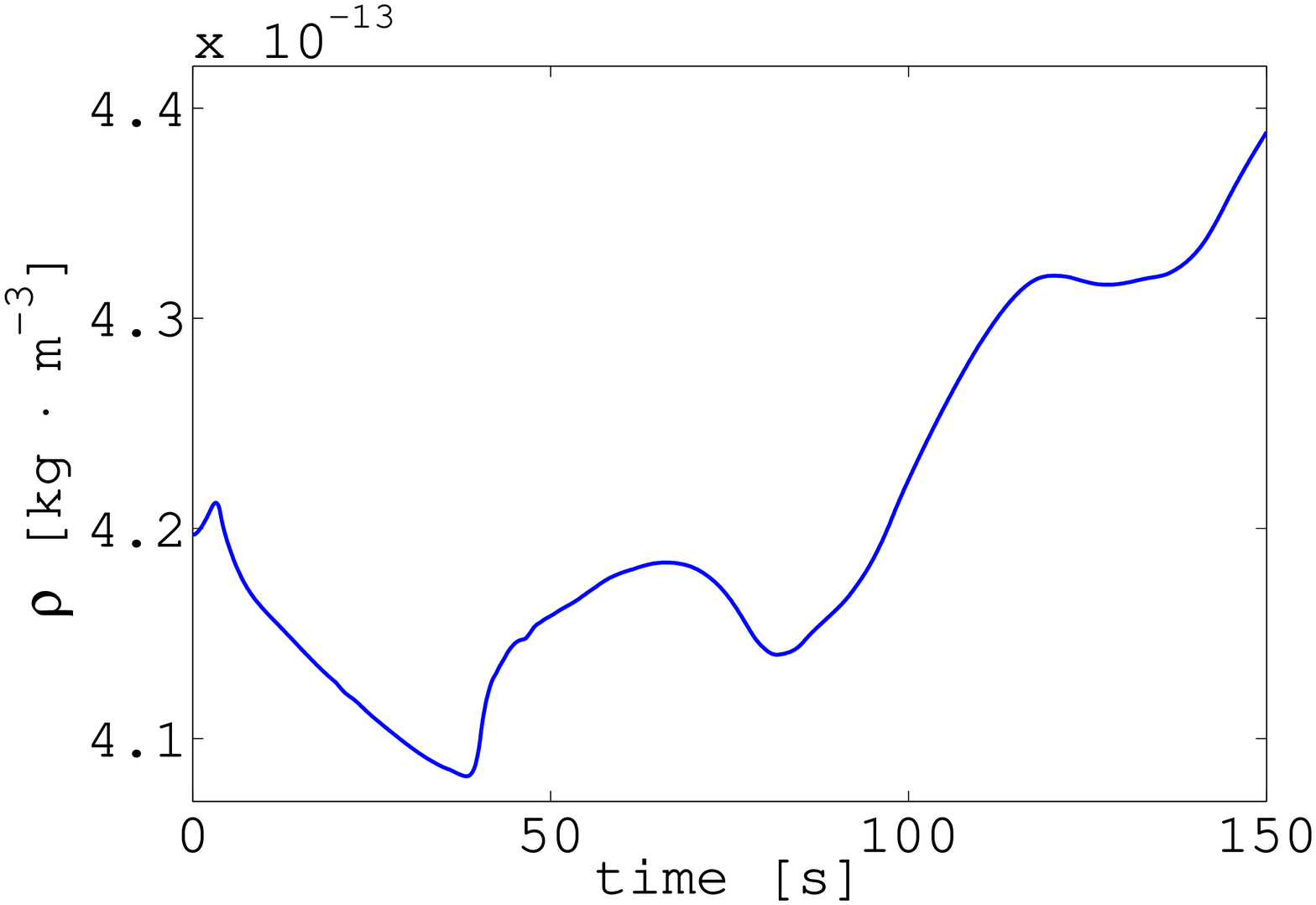}
\includegraphics[scale = 0.233]{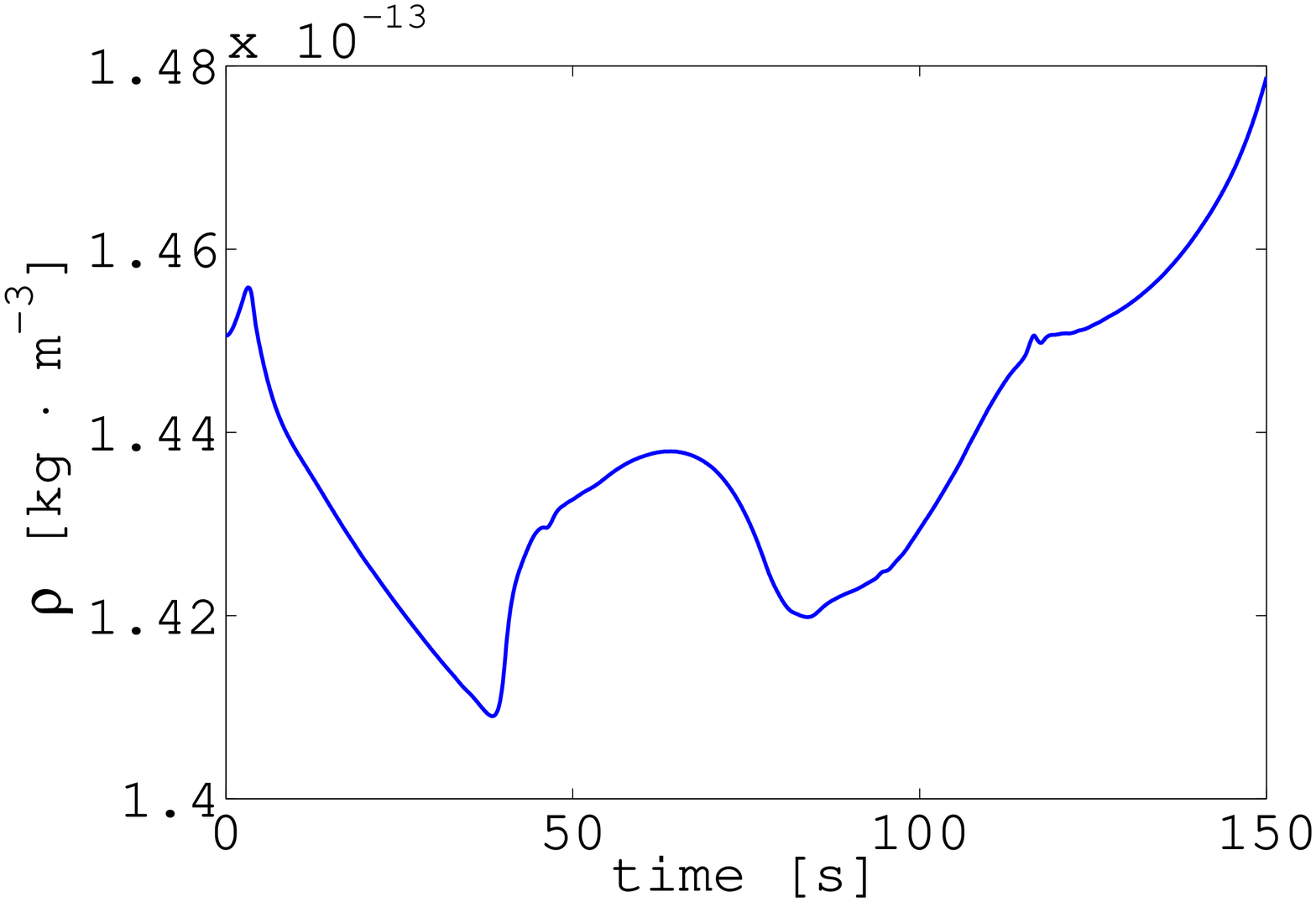}}
\mbox{
\includegraphics[scale = 0.233]{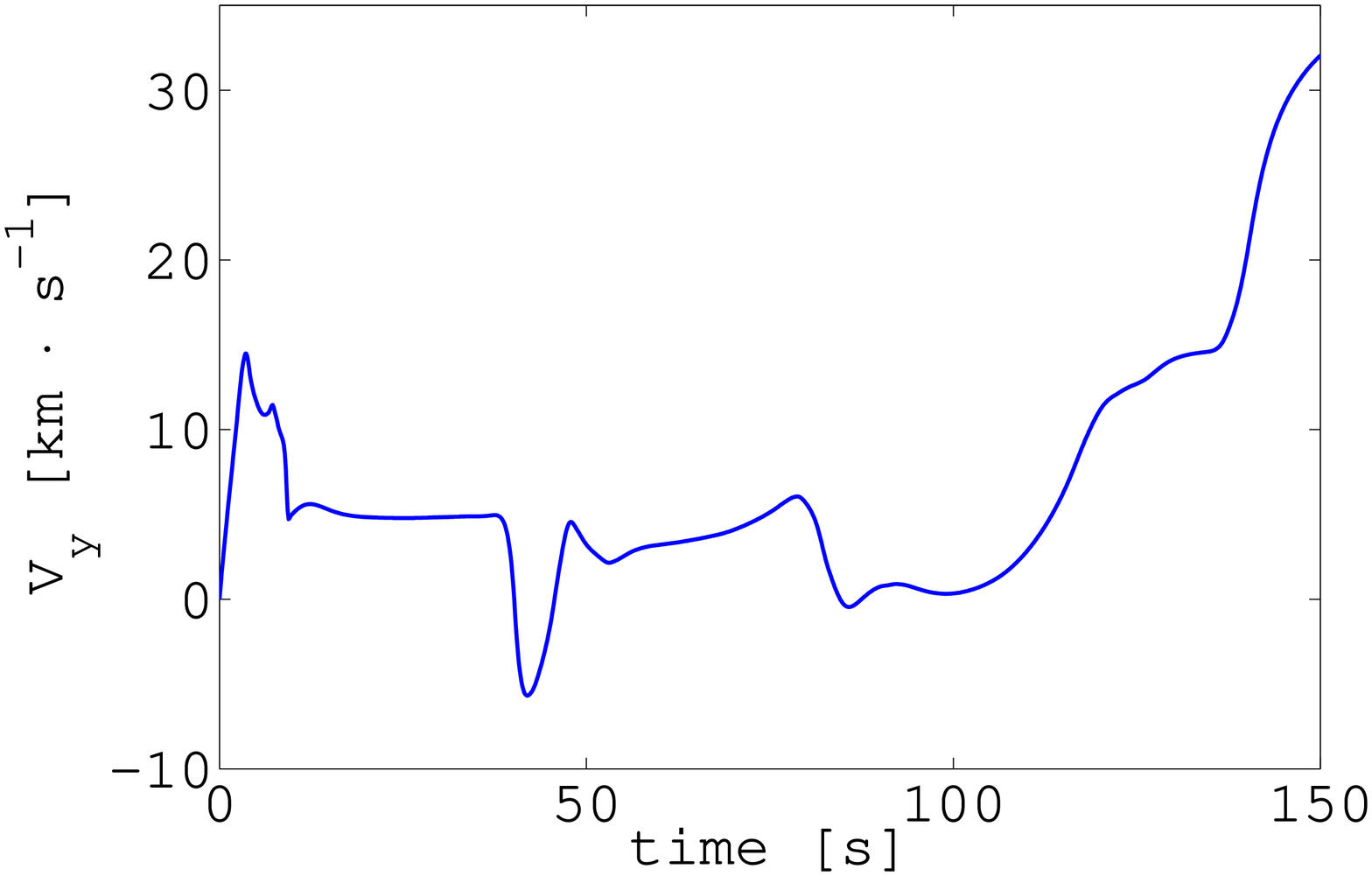}
\includegraphics[scale = 0.233]{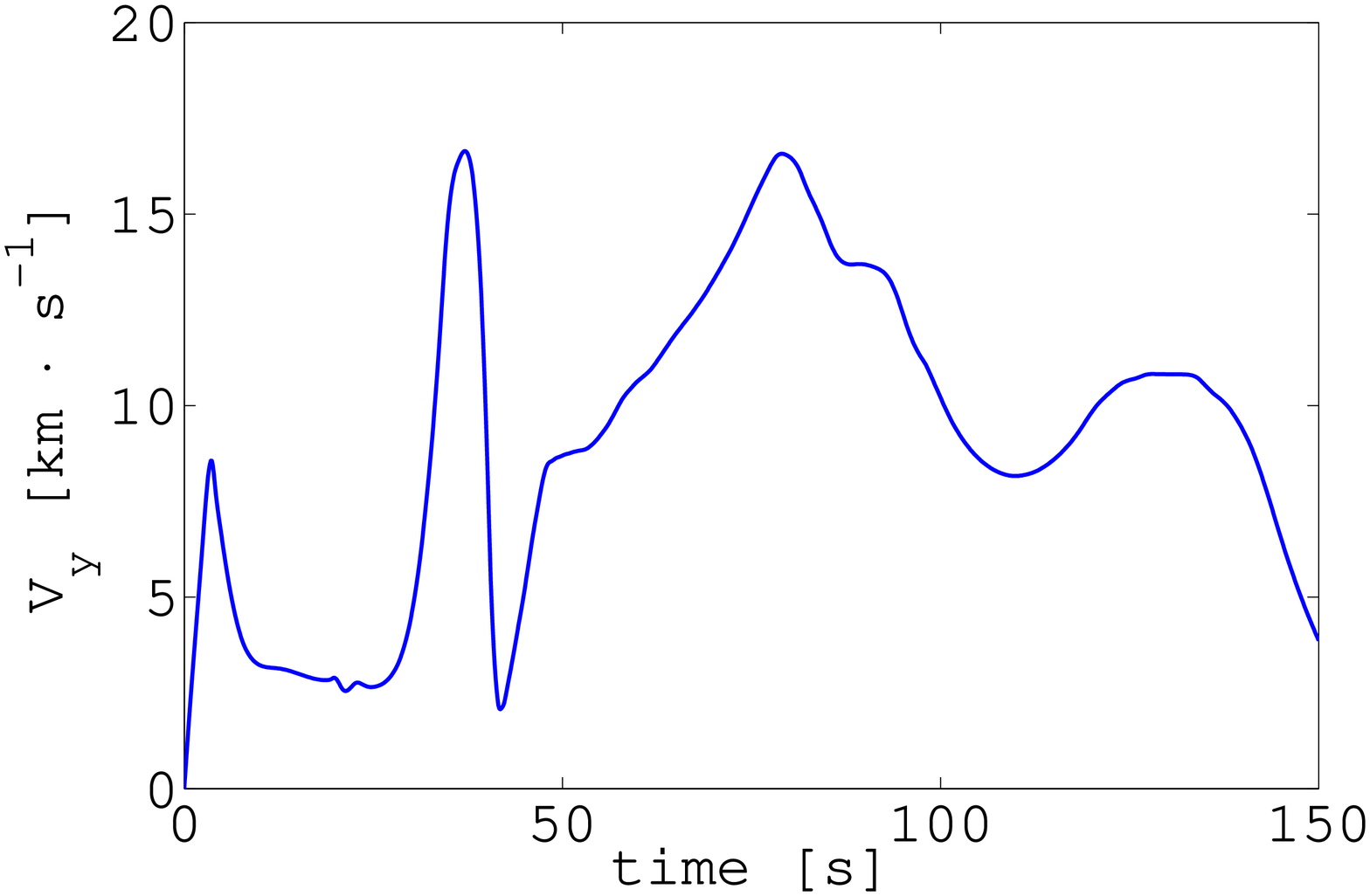}
\includegraphics[scale = 0.233]{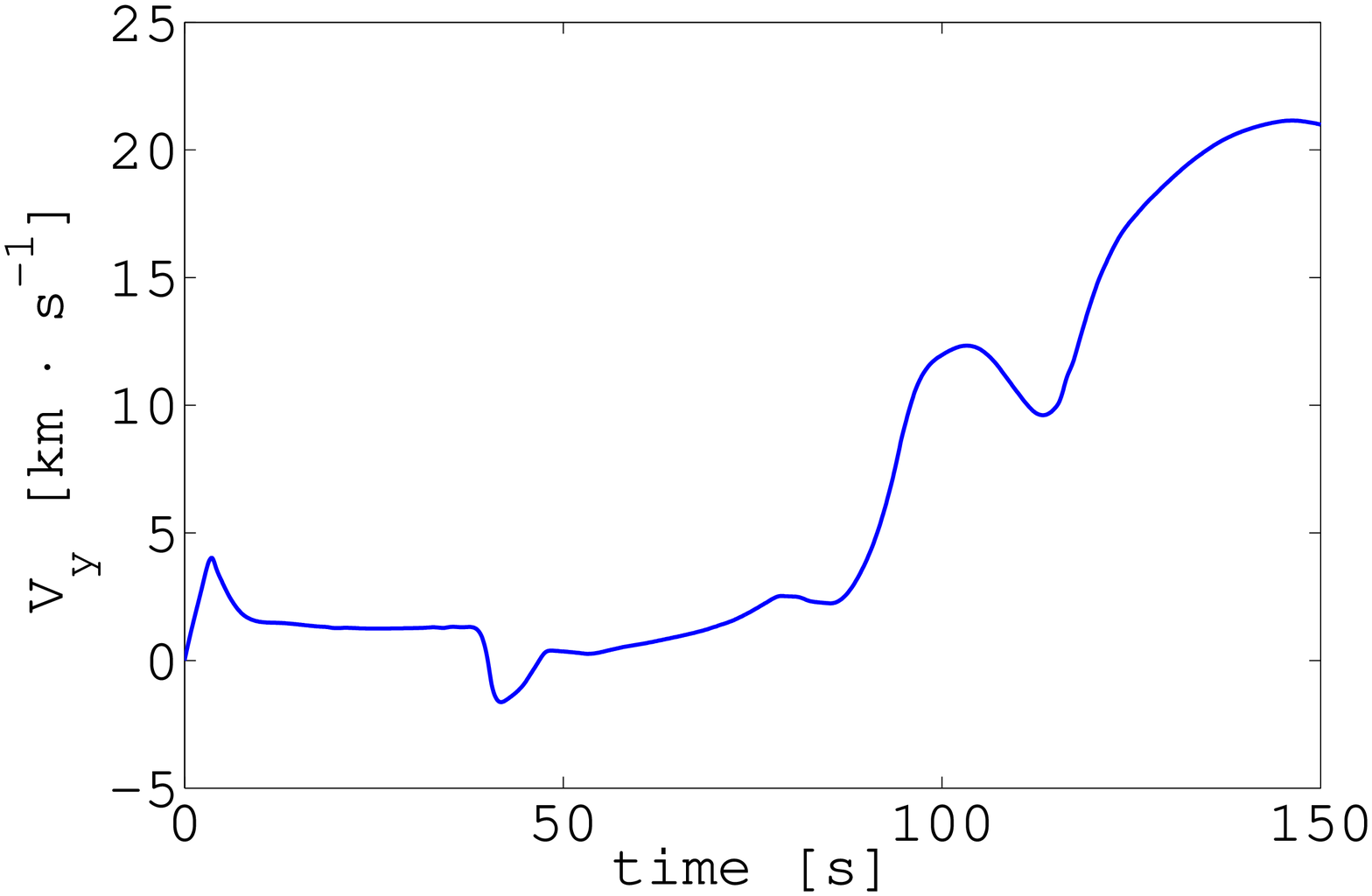}}
\mbox{
\includegraphics[scale = 0.233]{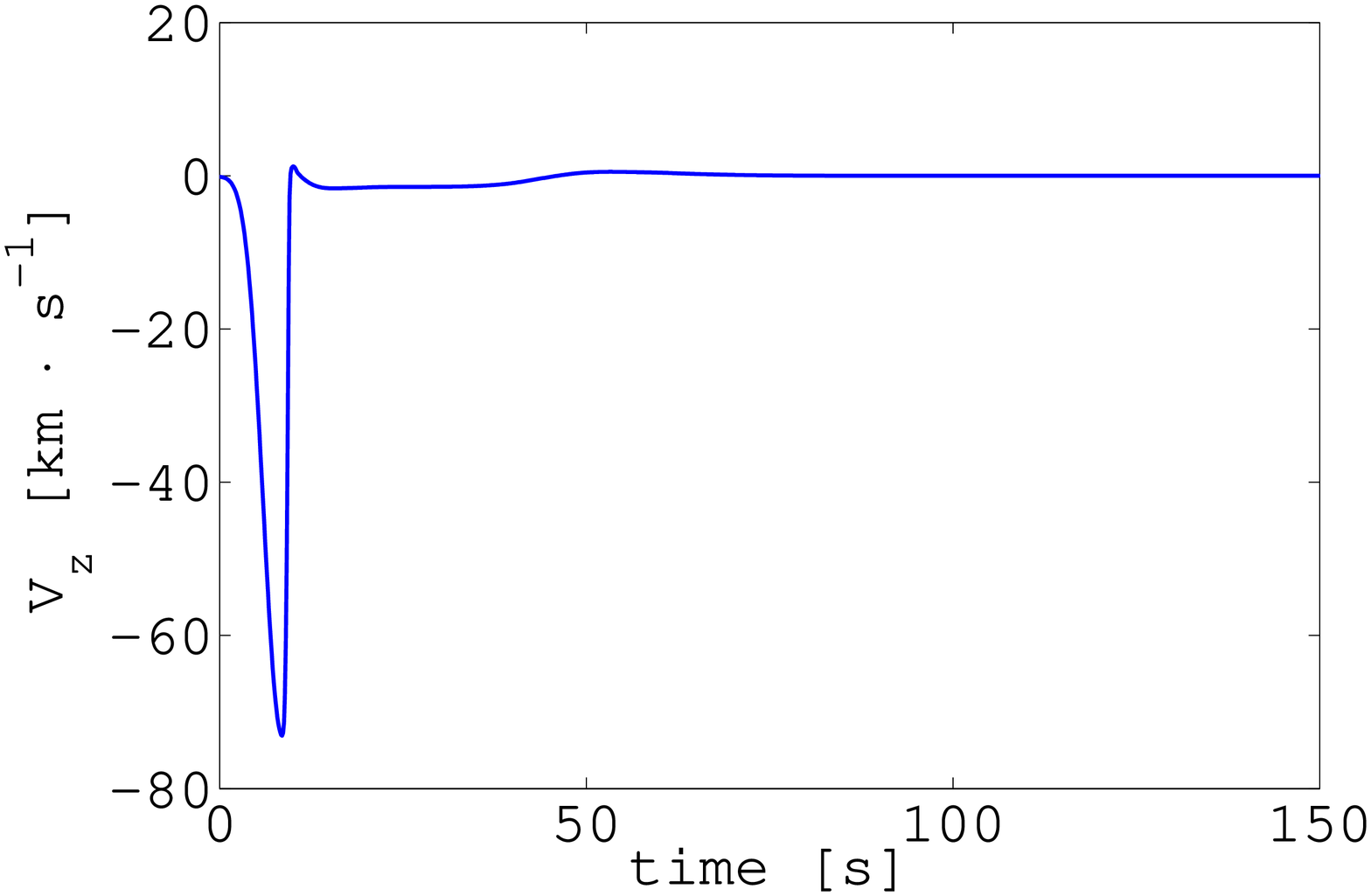}
\includegraphics[scale = 0.233]{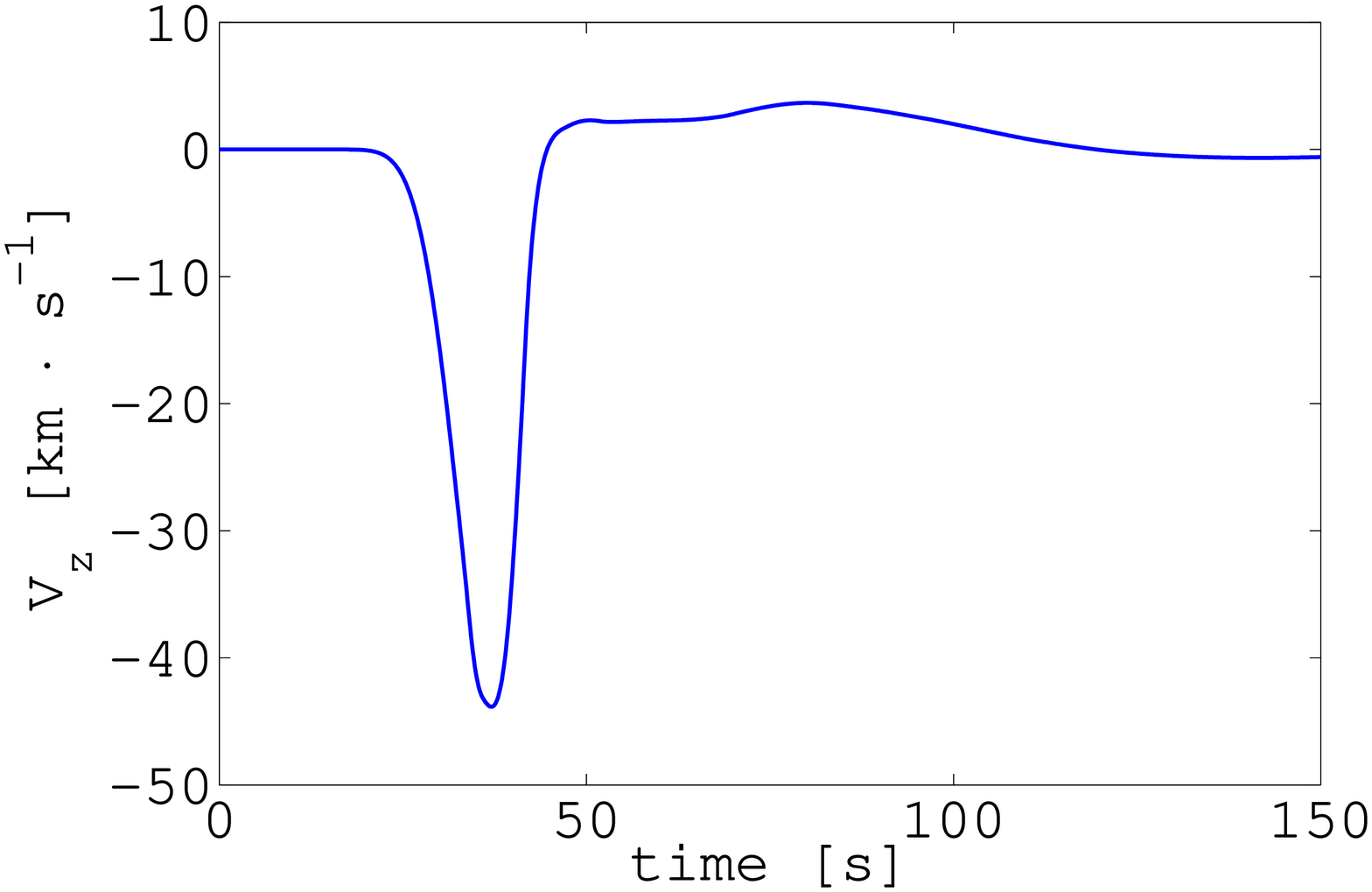}
\includegraphics[scale = 0.233]{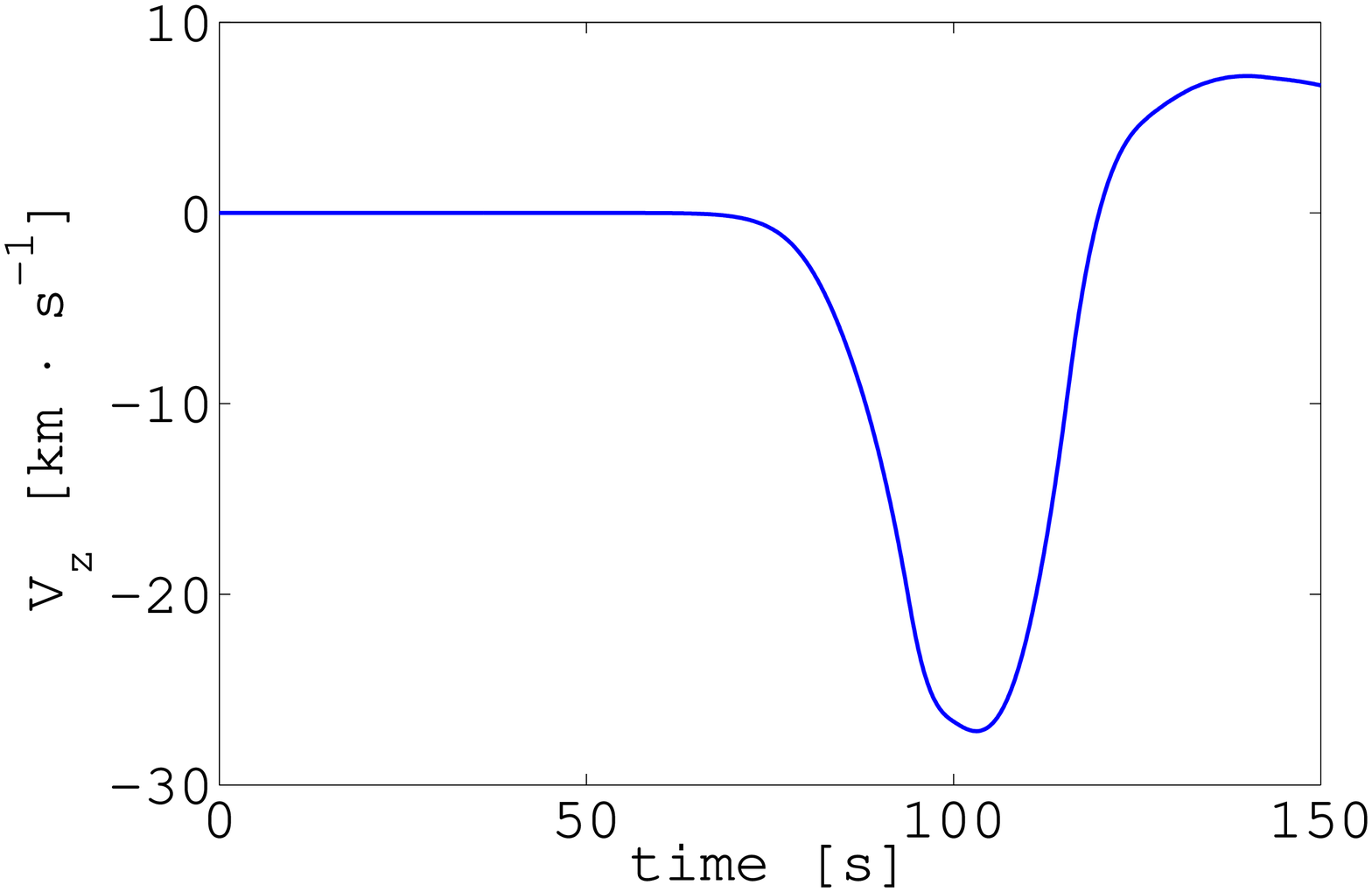}}

\caption{The top, middle and bottom panels display respectively the spatio-temporal averaged mass density $\varrho$, velocity components $V_y$ and $V_z$ at three different heights $15, 35$ and $55~\mathrm{Mm}$ above the solar surface.}
\label{fig304}
\end{figure*}


\subsection{Synthetic line-width increment due to Alfv\'en waves}

In the numerical domain, we have evolution of Alfv\'en waves in the Harris current-sheet, which propagate along open magnetic field lines in its stratified atmosphere. These Alfv\'en waves are described by the perturbations in the transverse velocity component ($V_{z}$). 
They result in the magneto-plasma as unresolved non-thermal motions (Banerjee et al. 1998; Chmielewski et al. 2013). The non-thermal motions due to Alfv\'en waves leave their signature in the line-profiles at each height in terms of their broadening. The increment in the line-width, therefore, provide the signature of Alfv\'en wave amplification (Banerjee et al. 1998, O'Shea et al. 2005). As mentioned above, the broadening of line-profiles due to Alfv\'en waves basically lead to the unresolved motions. This contribution to the spectrum, at any particular height in the 
solar atmosphere, can be due to various wave trains already passed through that detection point. However, these motions are not resolved by the spectrometer though inherent within the spectrum in terms of line-broadening. Therefore, we average over the space (entire pulse width), and time $(500~\mathrm{s})$ of various $V_{z}$ signals to know the resultant contribution of the unresolved non-thermal motions. The detailed discussions can be found in Chmielewski et al. (2013). In this sub-section, we describe the procedure that we adopt to convert the simulated velocity signals ($V_{z}$) into synthetic line-width of Fe XII $195.12~\mathrm{\AA}$. At any particular height in the model atmosphere, we average the wave velocity amplitude ($V_z$) over the entire pulse width between $t_{a}-250$ and $t_{a}+250~\mathrm{s}$, where $t_{a}$ is the arrival time of a wave signal to that particular height (Chmielewski et al. 2013). Therefore, the averaged transverse motions contribute to the non-thermal unresolved motions at a particular height in the model atmosphere as observed in the form of spectral line profile variation due to Alfv\'en waves. 
The averaged wave velocity amplitude is scaled in terms of non-thermal speed as $\xi^2 = 0.5\, V^2_{z}$  by taking into consideration the polarization and the direction of the propagation of Alfv\'en waves with respect to the line-of-sight (Banerjee et al. 1998). It should be noted that this scaling can be applicable for longer exposures. Our spatio-temporal averaging of velocity signal satisfies this requirement. Using the following formula (Mariska, 1992):
\beq
\label{eq:FWHM}
\sigma^{2} = \left[ 4\ln2\left(\frac{\lambda}{c}\right)^2 \left( \frac{2k_{\rm B}T}{m_{\rm i}}+\xi^2 \right)+{\sigma_{I}^{2}} \right], 
\eeq
we estimate the synthetic line-width (FWHM) of Fe XII ($\lambda = 195.12~\mathrm{\AA}$; $T_{f} = 1.2 \times 10^{6}~\mathrm{K}$) as a function of height in the model current-sheet where Alfv\'en waves are present. Here, $\sigma_{I}$ is the instrumental width associated with the Hinode/EIS spectrometer slit. The observed line-width is converted into FWHM and subtracted from instrumental width. While we derive the synthetic line-width of Fe XII $195.12~\mathrm{\AA}$ from numerical data, we ignore this factor as it only contains the physical information of thermal and non-thermal motions. 
It is clear from Fig.~7 (diamonds) that the synthetic line width resulting from the contribution of Alfv\'en waves grows with height, and its trend matches the variation of the observed line-width (cf., Fig.~2, bottom panel; beyond $10~\mathrm{Mm}$). 

We conclude that the Alfv\'en waves excited impulsively in our model atmosphere play a similar role like the transverse waves excited within the observed polar jet. Therefore, we exactly constrain the physical mechanisms, i.e., reconnection generated Alfv\'en waves, which also play a role in triggering the observed jet. The detailed comparison is given in the next sub-section. The width of the formed line-profile depends both on the thermal and non-thermal contributions in the emitted plasma. It is shown that the heating plays a crucial role in  the formation of spectral lines and its equivalent width (e.g. Peter et al. 2006). In the present observational base-line, we have found that our less intense jet does not show much temperature variations along its extended part in the corona as observed by Hinode/EIS. It is only clearly visible in Fe XII $195.12~\mathrm{\AA}$ line, while it is not evident in cooler (Si VII $275.35~\mathrm{\AA}$, $\log T_{f} = 5.0~\mathrm{MK}$) as well as hot (Fe XV $284.16~\mathrm{\AA}$, $\log T_{f} = 6.4~\mathrm{MK}$) lines. This means that the heating episode is quickly subsided after the reconnection generated origin of the jet. When it is scanned through Hinode/EIS 2$"$-slit, it has already reached its maximum height in the polar corona with the plasma maintained around inner coronal temperature of 1.0 MK. Therefore, the thermal width in the formed line profiles of Fe XII $195.12~\mathrm{\AA}$ (both observed and spectral) can be considered as a constant width equivalent to the formation temperature of Fe XII line, i.e., $T_{f} = 1.2 \times 10^{6}~\mathrm{K}$. The non-thermal contribution and resultant line broadening, therefore, result from the Alfv\'en wave amplification along the jet in its stratified atmosphere.

\begin{figure}
\centering
\includegraphics[width = 9.2cm, angle=0]{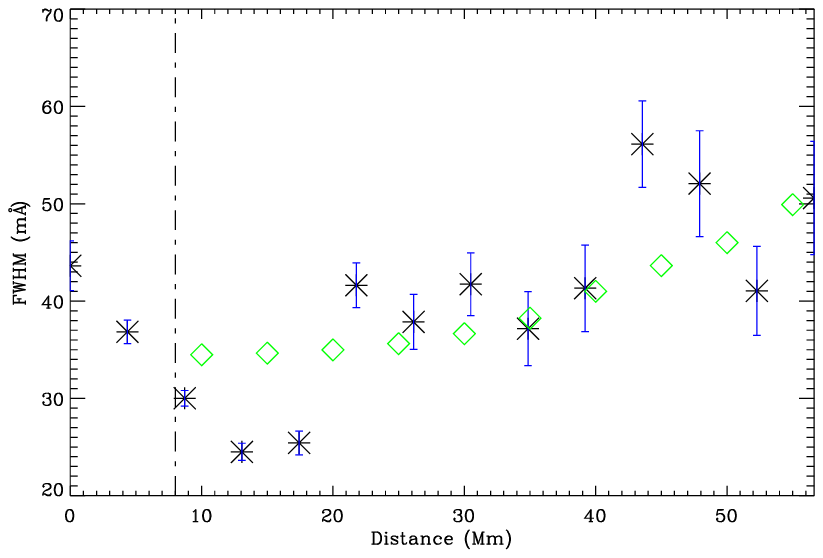}

\caption{The variation of synthetic line width with height (diamonds) of Fe XII $195.12~\mathrm{\AA}$ as derived by considering the contribution of propagating Alfv\'en waves in the Harris current-sheet, and the observed line-width broadening along the polar jet (asterisks).}
\label{fig304}
\end{figure}

\subsection{Comparison with the observational data}

The observed jet is triggered by reconnection between low-lying bipolar loops at its base, and pre-existing open field lines of the polar corona. This forms the typical inverted Y-shape configuration of the jet, and a vertical Harris current-sheet along the jet body. The X-type reconnection point can be created somewhere at coronal heights above the inverted Y-shape base  of the jet. We model this vertical Harris current-sheet as an upper part of the jet where waves and plasma perturbations occur. We assume that the magnetic reconnection generates a velocity pulse which results in Alfv\'en waves propagation along the coronal jet.

In the observations, we find the reconnection point at a height $5-10~\mathrm{Mm}$ from the base of the jet where maximum downflow speed is converted into blue-shift at higher heights along the jet. Up to this height, the measured FWHM is almost constant. It grows significantly beyond this height, and maximizes up to $50~\mathrm{Mm}$ height along the jet. As noted in Sec.~2, this provides the most likely signature of Alfv\'en wave growth along the jet body beyond the reconnection point. 

In our model, the transverse velocity perturbations are generated due to the reconnection in the lower part of the Harris current-sheet, which correspond to Alfv\'en waves. Although, we include only the reconnection generated velocity pulse in the vertical current-sheet of the jet. We obtain the time signatures of Alfv\'en waves at each height, and convert it into the unresolved non-thermal velocity. Since mass density decreases with height along the open field lines, Alfv\'en waves are amplified, contributing more to the non-thermal motions at higher altitudes. This results in the line width growth with height of any observed line profile in the stratified solar atmosphere. The synthesized FWHM (line-width) due to the contribution of Alfv\'en waves propagating in the current-sheet along the jet body. It also shows a similar increasing trend up to $y = 55~\mathrm{Mm}$ (cf., Fig.~7). It is constant up to $y = 15~\mathrm{Mm}$ above the reconnection point. Beyond the reconnection point, the observed FWHM profile along the polar jet (Fig.~3, bottom panel) increases. This confirms that the Alfv\'en waves are impulsively excited within the typical jet morphology due to magnetic reconnection. 
Alfv\'en waves propagate along the jet's open magnetic field lines, and further leave their signatures in the form of the observed line width  variation. Moreover, the excited slow magnetoaoustic waves perturb the plasma and result in the vertical flows along the magnetic field lines, which cause the jet plasma dynamics. In conclusion, our model supports the scenario of the evolution of non-linear Alfv\'en waves in the polar coronal jet, and associated flows that may contribute to its plasma evolution. 
However, it should be noted that we do not consider the jet's plasma evolution due to heating or direct ${\bf j} \times {\bf B}$ force that may also contribute initially in its formation.
\section{Discussions and conclusion}
Using the Hinode/EIS data, we observe that magnetic reconnection takes place at the coronal height within a polar jet. Above and below the reconnection point, the plasma shows upflows $(-7~\mathrm{km\,s}^{-1})$ and downflows $(+5~\mathrm{km\,s}^{-1})$, respectively. This confirms the formation of a typical reconnection-generated coronal jet. We also find the line-width increment (Fe XII $195.12~\mathrm{\AA}$) along the jet body beyond the reconnection point as a likely signature of the impulsively generated Alfv\'en waves. The observed line width increment with height matches with its theoretically estimated values, obtained by considering the contribution of Alfv\'en waves amplitude in the unresolved nonthermal motions of the plasma jet. The inference of this matching provides the physics of Alfv\'en waves which are driven within the coronal jets. The reconnection within the vertical Harris current-sheet along the jet's body trigger the longitudinal and transverse perturbation, which are associated with fast magnetoacoustic and Alfv\'en waves. The Alfv\'en waves leave their signatures collectively in the form of spectral line broadening with height, while the fast magnetoacoustic waves form the plasma flows in the jet. Linear and non-linear Alfv\'en waves can also be generated at the photosphere by granular motions. The region between photosphere and transition region acts as an Alfv\'enic resonator (e.g., Matsumoto \& Shibata 2010). In the present case, we do not find any observational signature of Alfv\'en waves below $10~\mathrm{Mm}$ within the jet. Therefore, we conjecture that the Alfv\'en waves are impulsively triggered above the reconnection site lying between $5-10~\mathrm{Mm}$ above the footpoint of the jet. The photospheric Alfv\'en waves generated by granular motions have typical initial amplitude of a few $\mathrm{km\,s}^{-1}$. When they propagate higher into the stratified atmosphere, they do amplify. However, in the present case, the Alfv\'en waves are triggered \textit{in situ} at coronal height and possess the larger instantaneous amplitudes ($>10~\mathrm{km\,s^{-1}}$). Therefore, we abbreviate them as impulsively generated non-linear Alfv\'en waves propagating along the polar jet.  

The underlying reconnection can drive Alfv\'en waves which excite the vertical plasma flow. We did not aim to understand the temporal evolution of jet plasma and its heating as previously explored by various models (e.g. Moreno-Insertis et al. 2013). Instead, we aim to understand the signature of Alfv\'en wave dynamics and associated vertical flows along the jet. We find the clues of both respectively in the form of line-width increment and Doppler velocity distribution (blue-shift) along the jet at various heights. For this purpose, we select the Hinode/EIS scanning spectral observations to understand the spatial variation of emission (flux) and plasma properties (flows and width) along the jet. The jet was initially launched due to reconnection which may also liberate the heat. The heat flux may be subsided during the whole half-life time of the jet when it has already reached a maximum altitude in the polar corona. The impulsive launch of the jet might leave other physical processes for rest of the time in space along with the jet, i.e., Alfv\'en waves and associated vertical flows. We have observed these two physical processes through EIS spectroscopic scan data,  which also confirm the model. Our model is simple in the sense that it does not include the radiative or thermal conductive losses. As a result, this model may not fully describe the reconnection generated heating and related additional plasma evolution within the jet as we do not include the thermodynamical losses in the governing energy equation. In general, the non-consideration of these terms may not affect the properties of Alfv\'en waves that evolved along the jet. However, they do affect the mass density and temperature evolution of the jet plasma. In our case, the jet is visible in the inner coronal temperature of 1.0 MK, and it is not a very hot evolution of the jet. Moreover, weak EUV emissions also suggest that plasma is not well evolved along the jet spine and it is not a bulky jet likewise X-ray jets and surges. In order to understand and compare the properties of the triggered Alfv\'en waves, we consider the simplified ideal situation in our model by ignoring the losses. These thermodynamical losses will be considered in our future works in which we will examine their effects on the evolution of temperature and density within the jet.

There exists only one report in the imaging X-ray observations from Hinode/XRT, which shows that the small amplitude Alfv\'en waves propagating from photosphere into the corona can drive the X-ray jets (Cirtain et al. 2007). Such important detections are debated by Van Doorsselaere et al. (2008a, 2008b). There are several attempts to model the polar jets in terms of the evolution of Alfv\'en waves (Kaghashvili 2009; Pariat et al. 2009, 2015; Hollweg \& Kaghashvili 2012, and references cited therein). These models are purely analytical/numerical without any observational support. Nevertheless, they typically reveal the physics of the jet forming regions as to how such waves can be excited within the typical magnetic field configurations of these jets. Nishizuka et al. (2008) have proposed a model of the coronal jets by extending the model of Yokoyama \& Shibata (1995), and found that direct reconnection generated forces can drive both the plasma jet and Alfv\'en waves at its base. Recently, Kayshap et al. (2013) reported that the reconnection in the lower atmosphere can generate the pulse train and associated slow shocks to drive the surge plasma.
Shibata et al. (2007) have also suggested three types of the processes for the jet formation
caused by reconnection at different heights. Depending upon the height of magnetic reconnection in the pre-existing magnetic field configuration of the jet, the hot X-ray, EUV, and cool jets as well as H$_{\alpha}$ surges can be triggered respectively in the coronal, transition region, and chromospheric reconnection. Therefore, reconnection can be an efficient mechanism to drive the waves/pulses within the confined magneto-plasma system (i.e., various jets) which can also play a role to power them. Alfv\'en waves are well studied both in theory and observations in driving the small-scale chromospheric jets (e.g., spicules; Kudoh \& Shibata 1999; McIntosh et al. 2011). 

In the present work, we have specifically made an effort to find the signatures of reconnction within a jet at a certain height from where the impulsively excited Alfv\'en waves and plasma motions are evident. We find the matching of the observed scenario with the given physical model of impulsive excitation of Alfv\'en waves within the vertical Harris current-sheet. Our model has the appropriate temperature conditions as well as stratified atmosphere in 2.5-D framework along with a consideration of the transition region which plays an important role in reflecting the Alfv\'en waves. The transition region has an important implication on contribution in non-thermal motions of the corona and in the formation of spectral lines provided they are excited below the transition region. In the present case, the site for impulsive energy release lies in the inner corona. Therefore, the excited Alfv\'en waves propagating in the outward direction will not have any influence on the transition region. Moreover, the appropriate temperature conditions also set the stratification within the model atmosphere, which will result in the appropriate growth of the Alfv\'en waves. Therefore, its contribution to the non-thermal motions and corressponding synthetic line broadening are adequately examined and studied. They also match with the observed line width variation along the polar jet. 

In conclusion, the numerical simulations show that the perturbations in the transversal component of velocity above the reconnection point within the current-sheet (the upper part of the jet and its reconnection site) can excite Alfv\'en waves impulsively. This can also trigger the plasma jet higher in the polar coronal hole as vertical plasma flows associated with magnetoacoustic waves. The inferred contribution of simulated Alfv\'en waves and their synthetic line widths match with the observed non-thermal line broadening along the jet. This provides a direct spectroscopic signature of the reconnection generated Alfv\'en waves within the polar jet in its typical magnetic field configuration (i.e., the Harris current-sheet and its X-type reconnection site).  

\label{obs}

\bigskip\noindent
{\bf Acknowledgments.}
{\bf The authors thank the anonymous referee for constructive comments}. AKS thanks the visiting scientist travel grant from Prof. K. Murawski, UMCS, Lublin, Poland during September-October 2014 where this work was initiated, and he also acknowledges Shobhna for all her supports. 
We acknowledge the use of Hinode/EIS observations in the present study. This work has granted an access to the HPC resources of CINES under the allocation 2012\ D0 046331 \& 2013\ D0046331 made by GENCI (Grand Equipement National de Calcul Intensif). PJ acknowledges support from the Grant P209/12/0103 of the Grant Agency of the Czech Republic. PJ and KM also thank the Marie Curie FP7-PIRSES-GA-2011-295272 Radiophysics of the Sun project. The authors also express their thanks to Piotr Chmielewski for his help with numerical data processing.
The FLASH code used in this work was in part developed by the DOE-supported ASC/Alliances Center for Astrophysical Thermonuclear Flashes at the University of Chicago.

%

\begin{thebibliography}{}
\bibitem[Banerjee et al.(1998)]{1998A&A...339..208B} Banerjee, D., Teriaca, L., Doyle, J.~G., Wilhelm, K.\ 1998, \aap, 339, 208
\bibitem[Bemporad \& Abbo(2012)]{2012ApJ...751..110B} Bemporad, A., Abbo, L.\ 2012, \apj, 751, 110
\bibitem[Chmielewski et al.(2013)]{2013MNRAS.428...40C} Chmielewski, P., Srivastava, A.~K., Murawski, K., Musielak, Z.~E.\ 2013, \mnras, 428, 40
\bibitem[Chmielewski et al.(2014)]{2014AcPPA.125..158C} Chmielewski, P., Srivastava, A.~K., Murawski, K., Musielak, Z.~E.\ 2014, AcPPA, 125, 158
\bibitem[Chung (2002)]{Chung} Chung, T. J. 2002, Computational Fluid Dynamics, Cambridge University Press, New York, USA
\bibitem[Cirtain et al.(2007)]{2007Sci...318.1580C} Cirtain, J.~W., Golub,L., Lundquist, L., et al.\ 2007, Science, 318, 1580 
\bibitem[Culhane et al.(2006)]{2006SPIE.6266E..0TC} Culhane, J.~L., Doschek, G.~A., Watanabe, T., et al.\ 2006, \procspie, 6266, 62660T
\bibitem[De Pontieu et al.(2007)]{2007Sci...318.1574D} De Pontieu, B., McIntosh, S.~W., Carlsson, M., et al.\ 2007, Science, 318, 1574 
\bibitem[Dwivedi \& Srivastava(2006)]{2006SoPh..237..143D} Dwivedi, B.~N., Srivastava, A.~K.\ 2006, \solphys, 237, 143
\bibitem[Dwivedi et al. (2014)]{2014PASJ..237..143D} Dwivedi, B.~N., Srivastava, A.~K., Mohan, A.\ 2014, PASJ, 66, S13 (1-11)
\bibitem[Filippov et al.(2009)]{2009SoPh..254..259F} Filippov, B., Golub, L., Koutchmy, S.\ 2009, \solphys, 254, 259
\bibitem[Fryxell et al. (2000)] {Fryxell} Fryxell, B. et al. 2000, ApJ Suppl. Ser. 131, 273
\bibitem[Galsgaard \& Roussev (2002)] {Galsgaard} Galsgaard, K., Roussev, I. 2002, \aap, 383, 685
\bibitem[Goossens et al.(2009)]{2009A&A...503..213G} Goossens, M., Terradas, J., Andries, J., Arregui, I., \& Ballester, J.~L.\ 2009, \aap, 503, 213
\bibitem[Goossens et al.(2012)]{2012ApJ...753..111G} Goossens, M., Andries, J., Soler, R., et al.\ 2012, \apj, 753, 111 
\bibitem[Harrison et al.(2002)]{2002A&A...392..319H} Harrison, R.~A., Hood, A.~W., Pike, C.~D.\ 2002, \aap, 392, 319
\bibitem[Hollweg \& Kaghashvili(2012)]{2012ApJ...744..114H} Hollweg, J.~V., Kaghashvili, E.~K.\ 2012, \apj, 744, 114 
\bibitem[Jess et al.(2009)]{2009Sci...323.1582J} Jess, D.~B., Mathioudakis, M., Erd{\'e}lyi, R., et al.\ 2009, Science, 323, 1582
\bibitem[Jel\'\i nek et al. (2012)]{jel4} Jel\'{\i}nek, P., Karlick\'y, M., Murawski, K. 2012, \aap, 546, A49
\bibitem[Kudoh \& Shibata(1999)]{1999ApJ...514..493K} Kudoh, T., Shibata, K.\ 1999, \apj, 514, 493
\bibitem[Lee \& Deane (2009)]{leeanddeane1}Lee, D., Deane, A. E. 2009, J. Comput. Phys., 228, 952
\bibitem[Lee (2013)]{lee1}Lee, D., 2013, J. Comput. Phys., 243, 269
\bibitem[Kayshap et al.(2013)]{2013ApJ...770L...3K} Kayshap, P., Srivastava, A.~K., Murawski, K., Tripathi, D.\ 2013, \apjl, 770, L3
\bibitem[Kayshap et al.(2013)]{2013ApJ...763...24K} Kayshap, P., Srivastava, A.~K., Murawski, K.\ 2013, \apj, 763, 24
\bibitem[Kaghashvili et al.(2009)]{2009ApJ...703.1318K} Kaghashvili, E.~K., Quinn, R.~A., Hollweg, J.~V.\ 2009, \apj, 703, 1318
\bibitem[Mariska(1992)]{1992str..book.....M} Mariska, J.~T.\ 1992, Cambridge Astrophysics Series, New York: Cambridge University Press
\bibitem[Mathioudakis et al.(2013)]{2013SSRv..175....1M} Mathioudakis, M., Jess, D.~B., Erd{\'e}lyi, R.\ 2013, \ssr, 175, 1
\bibitem[Matsumoto \& Shibata(2010)]{2010ApJ...710.1857M} Matsumoto, T., Shibata, K.\ 2010, \apj, 710, 1857
\bibitem[McIntosh et al.(2011)]{2011Natur.475..477M} McIntosh, S.~W., de Pontieu, B., Carlsson, M., et al.\ 2011, \nat, 475, 477
\bibitem[Moreno-Insertis \& Galsgaard(2013)]{2013ApJ...771...20M} Moreno-Insertis, F., Galsgaard, K.\ 2013, \apj, 771, 20
\bibitem[Murawski (2002)]{Murawski2} Murawski, K. 2002, Analytical and Numerical Methods for Wave Propagation in Fluid Media, World Scientific, Singapore
{\bf \bibitem[Murawski et al.(2015)]{2015A&A...577A.126M} Murawski, K., Solov'ev, A., Musielak, Z.~E., Srivastava, A.~K., \& Kra{\'s}kiewicz, J.\ 2015a, \aap, 577, A126}
{\bf \bibitem[Murawski et al.(2015)]{2015arXiv150503793M} Murawski, K., Srivastava, A.~K., Musielak, Z.~E., \& Dwivedi, B.~N.\ 2015b, arXiv:1505.03793} 
\bibitem[Nakariakov et al. (2004)]{Nakariakov3} Nakariakov, V. M., Arber, T. D., Ault, C. E., Katsiyannis, A. C., Williams, D. R., Keenan, F. P. 2004, Mon. Not. R. Astron. Soc. 349, 705
\bibitem[Nakariakov et al. (2005)]{Nakariakov4} Nakariakov, V. M., Pascoe, D. J., Arber, T. D. 2005, Space Sci. Rev. 121, 115
\bibitem[Nishizuka et al.(2008)]{2008ApJ...683L..83N} Nishizuka, N., Shimizu, M., Nakamura, T., et al.\ 2008, \apjl, 683, L83
\bibitem[Nistic{\`o} et al.(2009)]{2009SoPh..259...87N} Nistic{\`o}, G., Bothmer, V., Patsourakos, S., Zimbardo, G.\ 2009, \solphys, 259, 87 
\bibitem[Nistic{\`o} \& Zimbardo (2012)]{nis2} Nistic{\`o}, G. Zimbardo, G.\ 2012, Adv. Sp. Res., 49, 408
\bibitem[Okamoto et al.(2007)]{2007Sci...318.1577O} Okamoto, T.~J., Tsuneta, S., Berger, T.~E., et al.\ 2007, Science, 318, 1577
\bibitem[O'Shea et al.(2005)]{2005A&A...436L..35O} O'Shea, E., Banerjee, D., Doyle, J.~G.\ 2005, \aap, 436, L35 
\bibitem[Pariat et al.(2009)]{2009ApJ...691...61P} Pariat, E., Antiochos, S.~K., DeVore, C.~R.\ 2009, \apj, 691, 61
\bibitem[Pariat et al.(2015)]{2015A&A...573A.130P} Pariat, E., Dalmasse, K., DeVore, C.~R., Antiochos, S.~K., \& Karpen, J.~T.\ 2015, \aap, 573, AA130
\bibitem[Pascoe et al.(2010)]{pas1} Pascoe, D.~J., Wright, A.~N., De Moortel, I.\ 2010, \apj, 711, 990
\bibitem[Pascoe et al.(2011)]{pas2} Pascoe, D.~J., Wright, A.~N., De Moortel, I.\ 2011, \apj, 731, 73
\bibitem[Peter \& Judge(1999)]{1999ApJ...522.1148P} Peter, H., Judge, P.~G.\ 1999, \apj, 522, 1148
\bibitem[Peter et al.(2006)]{2006ApJ...638.1086P} Peter, H., Gudiksen, B.~V., Nordlund, {\AA}.\ 2006, \apj, 638, 1086
\bibitem[Priest (1982)]{Priest} Priest, E. R. 1982, Solar Magnetohydrodynamics, D. Reidel Publishing Company, London England 
\bibitem[Shibata(1982)]{1982SoPh...81....9S} Shibata, K.\ 1982, \solphys, 81, 9
\bibitem[Shibata et al.(2007)]{2007Sci...318.1591S} Shibata, K., Nakamura, T., Matsumoto, T., et al.\ 2007, Science, 318, 1591 
\bibitem[Somov(1994)]{1994SSRv...70..161S} Somov, B.V.\ 1994, SSRv, 70, 161
\bibitem[Srivastava \& Murawski(2011)]{2011A&A...534A..62S} Srivastava, A.~K., Murawski, K.\ 2011, \aap, 534, A62
\bibitem[Srivastava \& Goossens(2013)]{2013ApJ...777...17S} Srivastava, A.~K., \& Goossens, M.\ 2013, \apj, 777, 17 
\bibitem[Tomczyk et al.(2007)]{2007Sci...317.1192T} Tomczyk, S., McIntosh, S.~W., Keil, S.~L., et al.\ 2007, Science, 317, 1192
\bibitem[Toro (2006)]{toro1} Toro, E.~F.\ 2006, Int. J. Numer. Meth. Fl., 52, 433
\bibitem[Van Doorsselaere et al.(2008a)]{vandoors1} Van Doorsselaere, T., Brady, C.~S., Verwichte, E., Nakariakov, V.~M. \ 2008a, \aap, 491, L9
\bibitem[Van Doorsselaere et al.(2008b)]{2008ApJ...676L..73V} Van Doorsselaere, T., Nakariakov, V.~M., Verwichte, E.\ 2008b, \apjl, 676, L73
\bibitem[Vernazza et al.(1981)]{ver1} Vernazza, J.~E., Avrett, E.~H., Loeser, R., 1981, \apj, 45, 635
\bibitem[Yokoyama \& Shibata(1995)]{1995Natur.375...42Y} Yokoyama, T., Shibata, K.\ 1995, \nat, 375, 42 
\bibitem[Young et al.(2009)]{2009A&A...495..587Y} Young, P.~R., Watanabe, T., Hara, H., Mariska, J.~T.\ 2009, \aap, 495, 587 
\end{thebibliography}




\end{document}